\documentclass[12pt]{article}

\RequirePackage{natbib}
\usepackage{amsmath}
\usepackage{amsfonts}
\usepackage{amssymb}
\usepackage{amsthm}
\usepackage{rotating}
\usepackage{makecell}
\usepackage{hyperref}
\hypersetup{
    colorlinks,
    citecolor=black,
    filecolor=black,
    linkcolor=black,
    urlcolor=black
}
\usepackage{tabularx}
\usepackage{ctable}
\usepackage[font=footnotesize,labelfont=bf]{caption}
\usepackage{multirow}
\usepackage{color}
\usepackage{graphicx}
\usepackage{adjustbox}
\usepackage{pdflscape}
\usepackage[caption=false]{subfig}
\newcolumntype{C}{>{\centering\arraybackslash}X}%

\theoremstyle{plain}
\newtheorem{thm}{Theorem}

\begin{document}

\title{\Large \bf A phylogenetic scan test on Dirichlet-tree multinomial model for microbiome data}
\author{Yunfan Tang\footnotemark[1]\ , Li Ma\footnotemark[2]\ ,\ and Dan L. Nicolae\footnotemark[1] \\
{\normalsize\emph{University of Chicago\footnotemark[1]\ \ and Duke University\footnotemark[2]}}}

\date{}
\maketitle

\begin{abstract}
In this paper we introduce the phylogenetic scan test (PhyloScan) for investigating cross-group differences in microbiome compositions using the Dirichlet-tree multinomial (DTM) model. DTM models the microbiome data through a cascade of independent local DMs on the internal nodes of the phylogenetic tree. Each of the local DMs captures the count distributions of a certain number of operational taxonomic units at a given resolution. Since distributional differences tend to occur in clusters along evolutionary lineages, we design a scan statistic over the phylogenetic tree to allow nodes to borrow signal strength from their parents and children. We also derive a formula to bound the tail probability of the scan statistic, and verify its accuracy through simulations. The PhyloScan procedure is applied to the American Gut dataset to identify taxa associated with diet habits. Empirical studies performed on this dataset show that PhyloScan achieves higher testing power in most cases.
% * <marlee1982@gmail.com> 2016-10-15T14:42:29.516Z:
% 
% > tree-Dirichlet-multinomial
% I like PhyloDM slightly better than PhyloDM. It's less general but gives a more biological connotation.
% 
% ^ <yunfant@gmail.com> 2016-10-18T16:33:32.098Z.
% * <marlee1982@gmail.com> 2016-10-15T14:28:26.363Z:
%
% > p-value
%
% tail probability
%
% ^ <yunfant@gmail.com> 2016-10-18T16:05:50.750Z.
% * <marlee1982@gmail.com> 2016-10-15T14:27:20.639Z:
%
% > a scan statistic
%
% --> ``a scan statistic over the phylogenetic tree''
%
% ^ <yunfant@gmail.com> 2016-10-17T21:00:05.048Z.
% * <marlee1982@gmail.com> 2016-10-15T14:25:44.839Z:
%
% > adjacent nodes to borrow signal strength from each other
%
% --> ``nodes to borrow signal strength from their parents and children''
%
% ^ <yunfant@gmail.com> 2016-10-17T21:00:21.729Z.
% * <marlee1982@gmail.com> 2016-10-15T14:24:04.368Z:
%
% > of neighboring nodes
%
% --> ``along evolutionary lineages''
%
% ^ <yunfant@gmail.com> 2016-10-17T21:00:44.686Z.
\end{abstract}

\section{Introduction} \label{sec:Introduction}
Microbiome refers to the full collection of genes of all microbes in a community; for example, all bacteria in a sample from the gut of a healthy individual. The advent of next generation sequencing technologies, such as  Illumina Solexa, has allowed researchers to investigate the microbiome communities at an unprecedented level of quantification. The focus of this paper is on targeted amplicon sequencing and not on metagenome, but the ideas introduced here can be easily extended to metagenome data. A typical analysis pipeline involves sequencing one or a few of the variable regions of 16s ribosomal RNA, clustering the sequences into operational taxonomic units (OTU), assigning taxonomy to OTUs according to a reference database, and constructing a phylogenetic tree [e.g. \cite{Caporaso}]. There have been burgeoning efforts devoted to the study of human microbiome in the past decade, many of which aim at establishing evidence between microbiome and treatment effects or environmental covariates. Examples of such include associating gut microbiome with diet [\cite{David}], autism spectrum disorder [\cite{McDonald ASD}] and hormones [\cite{Neuman}]. Besides, a number of these studies are very large-scale initiatives such as Human Microbiome Project [\cite{HMP}] and American Gut [\cite{McDonald}], providing a broader understanding of the microbial variability. These studies jointly point to the fact that microbiome plays an integral part to our health, and much still remains to be explored in this area.
% * <marlee1982@gmail.com> 2016-10-15T14:29:57.271Z:
%
% > aims
%
% ``aim''
%
% ^ <yunfant@gmail.com> 2016-10-17T21:01:06.293Z.

The vast improvement in experimental tools contrasts with the slower development of statistical methods to analyze microbiome data. Typically, the majority of taxa can only be observed in a very small subset of samples, which causes the data table to be highly sparse. In addition, the within-group heterogeneity among samples leads to pronounced overdispersion in taxa proportions. Since standard multinomial distributions fail at capturing these features, Dirichlet-multinomial (DM) has been used as a natural extension. DM was originally proposed by \cite{Mosimann} and introduced into the microbiome context by \cite{Rosa} and \cite{Holmes}. Applying DM to test cross-group variation suffers from a number of drawbacks such as inability to localize any signal to a subgroup of taxa and reduced test power when a large number of taxa is present. Recent efforts to tackle these issues focus on incorporating phylogenetic tree into the model [\cite{Tang}, \cite{Silverman} and \cite{Wang and Zhao}]. In particular, \cite{Wang and Zhao} applied an extension of DM, namely Dirichlet-tree multinomial (DTM), first proposed by \cite{Dennis} under the name hyper-Dirichlet type 1 distribution. DTM is based on a decomposition of the sample space through a cascade of nested partitions similar to a Polya tree process [\cite{Lavine}]. Instead of placing a single global DM on all taxa, DTM consists of a collection of independent local DMs, each corresponding to a particular internal node on the phylogenetic tree. Since descendants of each internal node on the phylogenetic tree share a certain degree of evolutionary affinity, such decomposition strategy allows one to assign meaningful interpretation to each of the local DM distributions. An additional benefit is that the local DMs only target particular groups of taxa and consequently enjoy much lower degrees of freedom. This breaking down of the global distribution on all taxa counts allows testing each branch of the phylogeny individually, hence locating the signals to a certain taxonomic rank. The global cross-group test is therefore represented by a number of independent and biologically relevant constituents. For more general application of the Polya tree decomposition to hypothesis testing, see \cite{Ma and Wong}, \cite{Chen and Hanson}, \cite{Holmes2015} and \cite{Soriano and Ma}.

Although standard multiple testing procedures could be applied to results from testing all nodes, it is usually not the best practice to treat each hypothesis as a segregated entity. \cite{Soriano and Ma} pointed out that cross-group distributional variations tend to cluster, which causes hypotheses defined on nearby and/or nested windows more likely to be jointly true or false. This observation also holds in the microbiome data; cross-group differences in a certain ancestor node are more frequently accompanied with similar differences in its descendants. 
% * <marlee1982@gmail.com> 2016-10-15T14:46:01.492Z:
%
% > insightful
%
% remove ``insightful''? I don't want to call my own observation ``insightful''
%
% ^ <yunfant@gmail.com> 2016-10-17T21:06:09.610Z.
% * <marlee1982@gmail.com> 2016-10-13T16:02:09.327Z:
%
% > due to their functional similarity
%
% Remove ``due to their functional similarity''. This is not true---the clustering is often just a mathematical fact.
%
% ^ <yunfant@gmail.com> 2016-10-17T21:24:22.432Z:
%
% I think "functional similarity" gives a more specific reason why this pattern exists on microbiome data?
%
% ^ <yunfant@gmail.com> 2016-10-18T16:10:10.703Z.
To take advantage of this structure and optimize test power,  we adopt ideas from scan tests through constructing a collection of triplet statistics, each incorporating evidence from an internal node on the phylogenetic tree along with its parent and one of its children.
% * <marlee1982@gmail.com> 2016-10-13T16:04:04.933Z:
%
% 
% > we construct a tuple that consists of its two neighbors and the node itself
% we adopt ideas from scan tests through constructing a collection of triplet scan statistics each incorporates evidence from an internal node in the phylogenetic tree along with its parent and a child node.
% 
% ^ <yunfant@gmail.com> 2016-10-17T21:15:04.049Z.
%we adopt a strength-borrowing approach that shares the same spirit of \cite{Soriano and Ma}. For each internal node, 
% * <marlee1982@gmail.com> 2016-10-13T15:52:37.828Z:
%
% > %we adopt a strength-borrowing approach that shares the same spirit of \cite{Soriano and Ma}. For each internal node, 
%
% I would remove this because the borrowing strength approach used here is very different from the Markov tree used in Soriano and Ma (2016). In recent investigations, we found that in microbiome studies the clustering is mainly along lineages, and not shared among siblings, as such Markov tree is not the appropriate dependency structure for microbiome data. Instead, my group is writing a paper that use autoregressed logistic models to induce the ``chaining pattern''.
%
% ^ <yunfant@gmail.com> 2016-10-17T21:15:11.468Z.
The maximum of all these triplet statistics is used to test the global null hypothesis. Since the exact distribution of maximum statistic is intractable, we derive an upper and lower bound on its tail probability based on existing results on union probability [e.g., \cite{Hunter}, \cite{Efron} and \cite{Taylor}]. Our improved strategy first finds a subset consisting of independent components from the union, followed by bounding the probability of remaining components conditioned on the complement of that subset. A decay rate of the relative error of our approximation is also provided.
% * <marlee1982@gmail.com> 2016-10-15T14:42:00.903Z:
%
% > consisted
%
% consisting
%
% ^ <yunfant@gmail.com> 2016-10-17T21:19:23.312Z.
% * <marlee1982@gmail.com> 2016-10-13T16:03:26.448Z:
%
% > tuple
%
% change ``tuple'' to ``triplet'' across paper?
%
% ^ <yunfant@gmail.com> 2016-10-17T21:19:48.506Z.
% * <marlee1982@gmail.com> 2016-10-13T15:59:24.597Z:
% 
% > we construct a tuple that consists of its two neighbors and the node itself
% we adopt ideas from scan tests through constructing a collection of triplet scan statistics, each incorporating statistical evidence from an internal node in the phylogenetic tree along with that from its parent and a child node.
% 
% ^ <yunfant@gmail.com> 2016-10-17T21:19:50.186Z.

Section \ref{sec:Dirichlet-multinomial for microbiome data} briefly reviews the DM model. Section \ref{sec:Extension of Dirichlet-multinomial on a tree} formulates the DTM model and establishes its relation to the DM. Section \ref{sec:Scan statistic over the tree tuples} develops p-value approximation on the scan statistic for the DTM and verifies the result through simulation. Section \ref{sec:Application to American Gut dataset} applies the DTM model on the American Gut dataset to test the association of gut microbiome with a number of dietary habits. It also empirically demonstrates improvement of DTM over DM through likelihood ratio tests and comparing simulated test power. Section \ref{sec:Discussion} concludes with further discussions on potential DTM extensions.

\section{Dirichlet-multinomial for microbiome data} \label{sec:Dirichlet-multinomial for microbiome data} In this section we briefly recap the cross-group testing procedures on microbiome data using the Dirichlet-multinomial model, as presented in \cite{Rosa}.

Consider a microbiome dataset with $n$ samples and let $\Omega$ be the collection of a total of $K = |\Omega|$ OTUs. Without loss of generality, we assume $\Omega = \{1, 2, ..., K\}$. Each sample is a $K$-dimensional count vector representing the number of sequences in each of the $K$ OTUs. Let $\boldsymbol{x}_i = (x_{i1}, x_{i2}, ..., x_{iK})$ be the taxa count vector of the $i$th sample for $i = 1, 2,...,n$. In addition, define $N_{i.} = \sum_{j=1}^{K}x_{ij}$ to be the total number of sequences in the $i$th sample, $N_{.j} = \sum_{i=1}^{n}x_{ij}$ to be the total number of sequences in the $j$th OTU, and $N_{..} = \sum_{i=1}^n N_{i.} = \sum_{j=1}^K N_{.j}$. The Dirichlet-multinomial (DM) model assumes 
% * <marlee1982@gmail.com> 2016-10-15T14:47:55.198Z:
%
% > $i$th sample
%
% --> ``the $i$th sample
%
% ^ <yunfant@gmail.com> 2016-10-17T21:25:45.186Z.
% * <marlee1982@gmail.com> 2016-10-03T18:08:01.480Z:
%
% add ``the''
%
% ^ <yunfant@gmail.com> 2016-10-03T19:28:25.439Z.
% * <marlee1982@gmail.com> 2016-10-03T18:06:51.008Z:
%
% > $K = |\Omega|$ OTUs
%
% --> ``$K$ OTUs''
%
% ^ <yunfant@gmail.com> 2016-10-03T19:28:26.797Z.

$$\boldsymbol{q}_i \overset{i.i.d.}{\sim} \text{Dir}(\nu\boldsymbol{\pi}), \hspace{3mm}\boldsymbol{x}_i | \boldsymbol{q}_i \sim \text{Multinomial}(N_{i.}, \boldsymbol{q}_i),$$
where $\boldsymbol{\pi}=(\pi_1,\pi_2,\ldots,\pi_K)$ satisfies $\sum_{j=1}^{K}\pi_j = 1$, $\pi_j > 0$ denotes the mean taxa proportions and $\nu > 0$ is a dispersion parameter that controls the level of variation across samples. Alternatively one may use $\theta = \frac{1}{1+\nu}$ to parametrize the dispersion so that $0 \leq \theta < 1$. Integrating out the $\boldsymbol{q}_i$ gives 
% * <marlee1982@gmail.com> 2016-10-15T14:49:54.803Z:
%
% > $0 \leq \theta \leq 1$. 
%
% $0 \leq \theta < 1$
%
% ^ <yunfant@gmail.com> 2016-10-17T21:27:20.944Z.
% * <marlee1982@gmail.com> 2016-10-15T14:48:50.586Z:
%
% > $\nu \geq 0$
%
% --> ``$\nu >0$''
%
% ^ <yunfant@gmail.com> 2016-10-17T21:27:31.892Z.
% * <marlee1982@gmail.com> 2016-10-03T18:12:46.326Z:
% 
% > $\boldsymbol{\pi}$
% --> $\boldsymbol{\pi}=(\pi_1,\pi_2,\ldots,\pi_K)$
% 
% ^ <yunfant@gmail.com> 2016-10-03T21:34:10.759Z.
\begin{equation} \label{DM1}
f(\boldsymbol{x}_i) = {N_{i.} \choose \boldsymbol{x}_i} \frac{ \Gamma(\nu) } {\Gamma(N_{i.}+\nu)} \prod\limits_{j=1}^{K} \frac{\Gamma(x_{ij} + \nu\pi_j)}{\Gamma(\nu\pi_j)}.
\end{equation}

Throughout this paper we use $f(\cdot)$ exclusively to denote the DM probability mass function. When $\nu = \infty$ ($\theta = 0$), the DM degenerates to the standard multinomial distribution. Smaller values of $\nu$ indicates larger degrees of overdispersion. Assuming $\boldsymbol{x}_i$'s are independent, the likelihood function is simply the product of probabilities over all samples:
% * <marlee1982@gmail.com> 2016-10-03T18:14:43.299Z:
%
% > DM
%
% --> ``the DM''
%
% ^ <yunfant@gmail.com> 2016-10-03T21:33:34.954Z.
\begin{equation} \label{DM2}
\mathcal{L(\boldsymbol{\pi}, \nu)} = \prod\limits_{i=1}^{n} \bigg[ {N_{i.} \choose \boldsymbol{x}_i} \frac{ \Gamma(\nu) } {\Gamma(N_{i.}+\nu)} \prod\limits_{j=1}^{K} \frac{\Gamma(x_{ij} + \nu\pi_j)}{\Gamma(\nu\pi_j)} \bigg]
\end{equation}

As is shown in \cite{Weir and Hill}, the method of moments (MoM) estimates of the mean proportion $\boldsymbol\pi$ and dispersion $\theta$ are respectively
$$\hat{\boldsymbol\pi}=(\hat{\pi}_1, \hat{\pi}_2, ..., \hat{\pi}_K) \text{ with } \hat{\pi}_j = N_{.j}/N_{..}$$
$$\hat{\theta} =  \frac{\sum\limits_{j=1}^K (S_j - G_j)}{\sum\limits_{j=1}^K \big(S_j+(N_c-1)G_j \big)},$$
where we have $N_c = \frac{N_{..}-(N_{..})^{-1}\sum_{i=1}^nN_{i.}^2}{n-1},S_j = \frac{\sum_{i=1}^n N_{i.}(\hat{\pi}_{ij}-\hat{\pi}_j)^2}{n-1}$, and $G_j = \frac{\sum_{i=1}^n N_{i.} \hat\pi_{ij}(1-\hat\pi_{ij})}{\sum_{i=1}^n (N_{i.}-1)}$ with $\hat\pi_{ij} = x_{ij}/N_{i.}$.

For hypothesis testing, suppose we collect $G$ groups and the $g$th group data is given by $\boldsymbol{x}_1^{(g)}, \boldsymbol{x}_2^{(g)}, ..., \boldsymbol{x}_{n_g}^{(g)}$ with $N^{(g)}_{i.} = \sum_{j=1}^K x^{(g)}_{ij}$ and $N^{(g)}_{..} = \sum_{i=1}^n N^{(g)}_{i,}$. Similarly we define the $g$th group parameters as $\boldsymbol{\pi}^{(g)}, \nu^{(g)}$ with $\theta^{(g)} = \frac{1}{1+\nu^{(g)}}$. We wish to test the equality of mean proportion across all groups:
$$H_0: \boldsymbol{\pi}^{(1)} = \boldsymbol{\pi}^{(2)} = ... = \boldsymbol{\pi}^{(G)} \text{ vs } H_a: \text{otherwise}$$

Let $\hat{\boldsymbol\pi}^{(g)}$ and $\hat\theta^{(g)}$ be the MoM estimates of $\boldsymbol{\pi}^{(g)}$ and $\theta^{(g)}$, respectively. The cross-group pooled estimate of $\boldsymbol\pi$ is $\hat{\boldsymbol\pi}^{(Pool)} = \sum_{g=1}^G \bar s_g \hat{\boldsymbol\pi}^{(g)}$ with:
$$\bar s_g = \frac{(N^{(g)}_{..})^2C(\hat\theta^{(g)}, N^{(g)}_{..})^{-1}}{\sum_{r=1}^G (N^{(r)}_{..})^2C(\hat\theta^{(r)}, N^{(r)}_{..})^{-1}},$$
where
$$C(\hat\theta^{(g)}, N^{(g)}_{..}) = \hat\theta^{(g)} \Big(\sum_{i=1}^{n_g} (N^{(g)}_{i.})^2  -N^{(g)}_{..} \Big) + N^{(g)}_{..}.$$

Finally, the test statistic is defined as
\begin{equation} \label{DMtesting}
T = \sum_{g=1}^G (\hat{\boldsymbol\pi}^{(g)} - \hat{\boldsymbol\pi}^{(Pool)})^T (\bar S_g)^{-1} (\hat{\boldsymbol\pi}^{(g)} - \hat{\boldsymbol\pi}^{(Pool)}),
\end{equation}
where $\bar S_g$ is a diagonal matrix given by
$$\bar S_g = \Big( (N^{(g)}_{..})^2 C(\hat\theta_g,N^{(g)}_{..})^{-1}  \Big)^{-1} D(\hat{\boldsymbol\pi}^{(Pool)}),$$
and $D(\hat{\boldsymbol\pi}^{(Pool)})$ is also diagonal with diagonal elements given by $\hat{\boldsymbol\pi}^{(Pool)}$. The asymptotic distribution of $T$ under $H_0$ is $\chi^2_{(K-1)(G-1)}$ as $n_g \rightarrow \infty$ for all $g$.

\section{Dirichlet-tree multinomial and hypothesis testing} \label{sec:Extension of Dirichlet-multinomial on a tree}
In order to incorporate the phylogenetic tree into the model, \cite{Wang and Zhao} considered an extension named Dirichlet-tree multinomial (DTM). DTM allows us to separately test cross-group differences in each internal node, locating the source of overall difference within particular subgroups of OTUs. Each of the local test, by design, has the benefit of reduced degrees of freedom.

\subsection{Model formulation} \label{sec:Extension of Dirichlet-multinomial on a tree-Model formulation}

Let $\mathcal{T} = (\Omega, \mathcal{I})$ be a rooted phylogenetic tree where the set of OTUs $\Omega$ are placed on the leaves and $\mathcal{I}$ is the set of all internal nodes. We represent the elements in $\mathcal{I}$ to be subsets of $\Omega$ since each internal node is uniquely identified by the subset of all OTUs underneath it, and vice versa. Each subset of OTU that corresponds to an internal node shares a hypothetical ancestor along the lineage. Additionally, each leaf node is uniquely identified by a singleton set consisting of that particular OTU.
% * <marlee1982@gmail.com> 2016-10-15T14:56:07.391Z:
%
% > such
%
% --> ``a''
%
% ^ <yunfant@gmail.com> 2016-10-17T21:36:44.428Z.
% * <marlee1982@gmail.com> 2016-10-15T14:54:20.482Z:
%
% > internal 
%
% internal node
%
% ^ <yunfant@gmail.com> 2016-10-17T21:36:46.231Z.

\begin{figure}[h]
\centering\includegraphics[height=7cm]{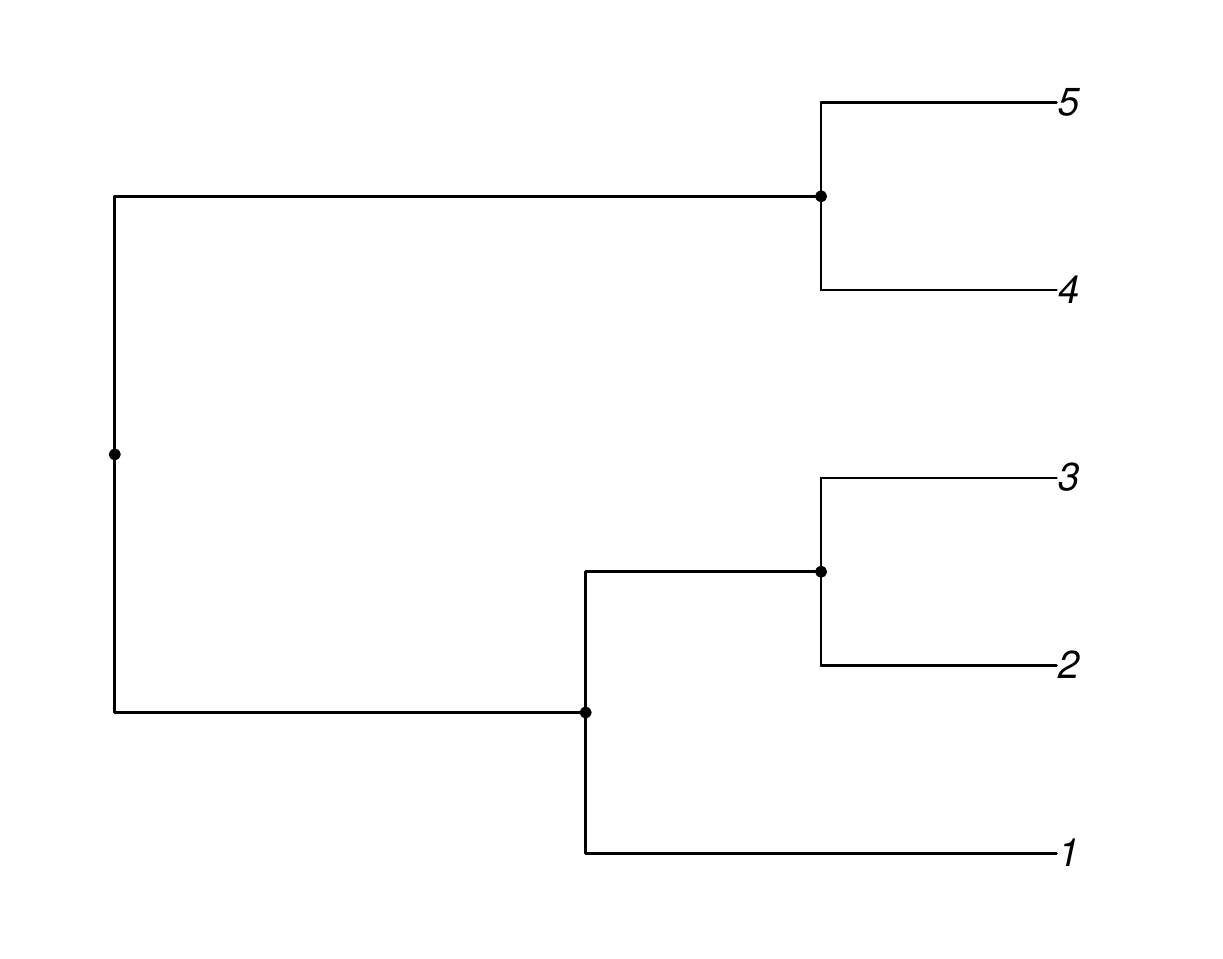}
\caption{\label{fig:Phylotree example} An example of a phylogenetic tree with five OTUs.}
\end{figure}

Figure \ref{fig:Phylotree example} shows an example of a simple phylogenetic tree over 5 OTUs and 4 internal nodes. This tree has $\Omega = \{1,2,3,4,5\}$ and $\mathcal{I} = \big\{ \{1,2,3,4,5\}, $ $\{1,2,3\}, \{4,5\}, \{2,3\} \big\}$.

Now for $\forall A \in \mathcal{I}$, let $\mathcal{C}(A)$ be the collection of $A$'s child nodes in $\mathcal{T}$. The elements of $\mathcal{C}(A)$ are also subsets of $\Omega$. Also $\forall A \in \mathcal{I} \cup \big\{ \{\omega\} | \omega \in \Omega\big\}, A \neq \Omega$, let $R(A)$ denote the parent node of $A$. In Figure 1, for example, $\mathcal{C}(\{1,2,3\}) = \big\{ \{1\}, \{2,3\}\big\}$ and $R(\{1,2,3\}) = \{1,2,3,4,5\} = \Omega$. Notice that certain $\mathcal{C}(A)$'s contain singletons of $\Omega$ since some children  are leaves. Let $k(A) = |\mathcal{C}(A)|$ be the number of children under $A$ and write $\mathcal{C}(A) = \{\mathcal{C}(A)_1, \mathcal{C}(A)_2, ..., \mathcal{C}(A)_{k(A)}\}$. For each $i = 1, 2, ..., n$ and $j = 1, 2, ..., k(A)$, let 
$$x_{ij}(A) = \sum\limits_{\omega \in \mathcal{C}(A)_j} x_{i\omega}$$
be the count of the $j$th child of $A$ in the $i$th sample. The count vector associated with $A$ is therefore
% * <marlee1982@gmail.com> 2016-10-15T14:57:13.787Z:
%
% > $i$th sample
%
% --> ``the $i$th sample''
%
% ^ <yunfant@gmail.com> 2016-10-17T23:17:54.497Z.
$$\boldsymbol{x}_i(A) = \big(x_{i1}(A), x_{i2}(A), ..., x_{ik(A)}(A)\big) $$
with the sum $N_i(A) = \sum_{j=1}^{k(A)}x_{ij}(A) = \sum_{\omega \in A}x_{i\omega}$. It is straightforward to see that $N_i\big(\mathcal{C}(A)_j\big) = x_{ij}(A)$ for $j = 1, 2, ..., k(A)$. In addition, we always have $N_i(\Omega) = N_{i.}$.

The DTM distribution separately models the count vector $\boldsymbol{x}_i(A)$ conditional on $N_i(A)$ for each $A$. Specifically for $\forall A \in \mathcal{I}$,
\begin{equation}\label{PhyloDM0}
\boldsymbol{q}_{A,i} \overset{i.i.d.}{\sim} \text{Dir}(\nu_A\boldsymbol{\pi_A}), \hspace{3mm} \boldsymbol{x}_i(A) | N_i(A), \boldsymbol{q}_{A,i} \sim \text{Multinomial}(N_i(A), \boldsymbol{q}_{A,i})
\end{equation}
% * <marlee1982@gmail.com> 2016-10-15T14:58:15.013Z:
%
% > \begin{equation}\label{PhyloDM0}
% > \boldsymbol{q}_{A,i} \overset{i.i.d.}{\sim} \text{Dir}(\nu_A\boldsymbol{\pi_A}), \boldsymbol{x}_i(A) | N_i(A), \boldsymbol{q}_{A,i} \sim \text{Multinomial}(N_i(A), \boldsymbol{q}_{A,i})
% > \end{equation}
%
% add more spacing between ``$N_i(A)$, '' and ``\boldsymbol{q}_{A_i}\sim''
%
% ^ <yunfant@gmail.com> 2016-10-17T21:39:09.171Z:
%
% I think you meant to add more spaces between "\text{Dir}(\nu_A\boldsymbol{\pi_A})" and "\boldsymbol{x}_i(A)", which I just did.
%
% ^ <yunfant@gmail.com> 2016-10-17T21:43:27.011Z:
%
% Adding more space between ``$N_i(A)$, '' and ``\boldsymbol{q}_{A_i}\sim'' might make it harder for readers to see that "\boldsymbol{x}_i(A)" are conditioned on both of them.
%
% ^ <marlee1982@gmail.com> 2016-10-18T17:26:42.801Z:
%
% yes exactly the original spacing is very confusing. it looks much better now, perhaps add a little more?
%
% ^ <yunfant@gmail.com> 2016-10-18T18:32:22.080Z.
where $\nu_A > 0$ is the overdispersion parameter of the counts of $A$'s children and $\boldsymbol{\pi}_A = (\pi_{A,1}, \pi_{A,2}, ..., \pi_{A,k(A)})$ satisfying $\sum_{i=1}^{k(A)}\pi_{A,i} = 1$ denotes their mean proportion. The Dirichlet prior distribution of all $A$'s are mutually independent. Integrating out $q_{A,i}$ gives
% * <marlee1982@gmail.com> 2016-10-15T14:59:17.308Z:
% 
% > are mutually independent
% --> ``be mutually independent''
% 
% ^ <yunfant@gmail.com> 2016-10-17T21:40:04.028Z.
\begin{equation} \label{PhyloDM1}
f\big(\boldsymbol{x}_i(A)|N_i(A)\big) = {N_i(A) \choose \boldsymbol{x}_i(A)} \frac{ \Gamma(\nu_A) } {\Gamma(N_i(A)+\nu_A)} \prod\limits_{j=1}^{k(A)} \frac{\Gamma(x_{ij}(A) + \nu_A\pi_{A,j})}{\Gamma(\nu_A\pi_{A,j})},
\end{equation}
which ultimately yields
\begin{equation} \label{PhyloDM2}
f_{T}(\boldsymbol{x}_i) = \prod\limits_{A \in \mathcal{I}} f\big(\boldsymbol{x}_i(A)|N_i(A)\big) 
\end{equation}
and likelihood function
\begin{equation} \label {PhyloDM3}
\mathcal{L}_T(\{(\nu_A, \boldsymbol{\pi}_A): A \in \mathcal{I} \}) = \prod\limits_{i=1}^n \prod\limits_{A \in \mathcal{I}} f\big(\boldsymbol{x}_i(A)|N_i(A)\big) 
\end{equation} 
with $f_T(\cdot)$ and $\mathcal{L}_T(\cdot)$ denoting the DTM probability mass function and likelihood function respectively. The representations in (\ref{PhyloDM1}) and (\ref{PhyloDM2}) naturally lead to a top-down generative scheme of the count data on the nodes, as each layer of DM models a subset of OTU counts at increased level of resolution conditioned on their sum.

Interestingly, \cite{Dennis} showed that the global DM distribution on OTU counts is nested in the DTM family. A simple explanation of this relation is that both the global Dirichlet prior and the multinomial probabilities can be factorized over $\mathcal{I}$, i.e.
\begin{equation*}
\boldsymbol{q}_i \sim \text{Dir}(\nu\boldsymbol{\pi}) \Leftrightarrow \forall A \in \mathcal{I}, \hspace{3mm} \frac{\boldsymbol{q}_i(A)}{\sum\limits_{\omega \in A}q_{i\omega} } \sim \text{Dir}\Big(\nu\sum\limits_{\omega \in A}\pi_{\omega} \cdot  \frac{\boldsymbol{\pi}(A)}{\sum\limits_{\omega \in A}\pi_{\omega} } \Big) \text{ independently} 
\end{equation*}
% * <marlee1982@gmail.com> 2016-10-15T15:01:05.715Z:
%
% > \begin{equation*}
% > \boldsymbol{q}_i \sim \text{Dir}(\nu\boldsymbol{\pi}) \Leftrightarrow \forall A \in \mathcal{I}, \frac{\boldsymbol{q}_i(A)}{\sum\limits_{\omega \in A}q_{i\omega} } \sim \text{Dir}\Big(\nu\sum\limits_{\omega \in A}\pi_{\omega} \cdot  \frac{\boldsymbol{\pi}(A)}{\sum\limits_{\omega \in A}\pi_{\omega} } \Big) \text{ independently} 
% > \end{equation*}
%
% more spacing after ``,''
%
% ^ <yunfant@gmail.com> 2016-10-17T23:18:32.852Z.

\begin{equation*}
\boldsymbol{x}_i | \boldsymbol{q}_i \sim \text{Multinomial}(N_{i.}, \boldsymbol{q}_i) \Leftrightarrow 
\end{equation*}
$$\forall A \in \mathcal{I}, \hspace{3mm}
\boldsymbol{x}_i(A) | \boldsymbol{q}_i, N_i(A) \sim  \text{Multinomial}(N_i(A), \frac{\boldsymbol{q}_i(A)}{\sum\limits_{\omega \in A}q_{i\omega} }) $$
where we similarly defined 
$$q_{ij}(A) = \sum\limits_{\omega \in \mathcal{C}(A)_j}q_{i\omega}, \boldsymbol{q}_i(A) = (q_{i1}(A), q_{i2}(A), ..., q_{ik(A)}(A)) $$
$$\pi_j(A) = \sum\limits_{\omega \in \mathcal{C}(A)_j}\pi_{\omega}, \boldsymbol{\pi}(A) = (\pi_1(A), \pi_2(A), ..., \pi_{k(A)}(A)) $$

In the DTM representation of global DM, the overdispersion and mean proportion of the counts of $A$'s children are respectively  $\nu_A = \nu\sum_{\omega\in A} \pi_{\omega}$ and $\boldsymbol{\pi}_A = \boldsymbol{\pi}(A)/\sum_{\omega \in A}\pi_\omega$. It is not hard to notice that there is a bijective correspondence between $\boldsymbol{\pi}$ and $\{\boldsymbol{\pi}_A  = \boldsymbol{\pi}(A)/\sum_{\omega \in A}\pi_\omega: A \in \mathcal{I} \}$. In addition, all of these local dispersions are governed by a single global $\nu$, which is highly restrictive as it does not allow any node-specific characterization of within-group variation. Section \ref{sec:Application to American Gut dataset-DM vs PhyloDM test} provides likelihood ratio test results supporting this claim.

\subsection{Hypothesis testing} \label{sec:Extension of Dirichlet-multinomial on a tree-Hypothesis testing}
The DTM model in (\ref{PhyloDM2}) and (\ref{PhyloDM3}) motivates a node-by-node testing strategy for cross-group comparison. To compare the proportion across G groups of observations, we carry out an MoM test using (\ref{DMtesting}) individually for each $A$, i.e.
$$H_{0,A}: \boldsymbol{\pi}^{(1)}_A = \boldsymbol{\pi}^{(2)}_A = ... = \boldsymbol{\pi}^{(G)}_A \text{ vs } H_{a,A}: \text{otherwise}$$

Each of the MoM test statistic are calculated conditional on $\{N_i^{(g)}(A)| 1\leq g\leq G, 1\leq i \leq n_g\}$, where $N_i^{(g)}(A)$ is the sum of OTU counts under $A$ in the $i$th sample of $g$th group. The test statistic for $H_{0,A}$ only has degrees of freedom $(G-1)(k(A)-1)$, much smaller than the degrees of freedom for DM test as $(G-1)(K-1)$. The local DM test is therefore more powerful than the global DM test, provided that the extent of cross-group difference on the internal nodes is not diluted too much as we group multiple OTUs together. Obviously, the extent of dilution is largely determined by the tree structure. The ideal scenario is that OTUs placed under the same internal node $A$ demonstrate increasing or decreasing abundance simultaneously for all samples in a certain group, so $H_{0,R(A)}$ will be most effective. This also motivates using the phylogenetic tree to carry out the decomposition, as functionally similar OTUs tend to exhibit similar abundance changes within the same group.
% * <marlee1982@gmail.com> 2016-10-15T15:07:40.766Z:
%
% > the effect size
%
% --> ``the extent of cross-group difference''
% because we didn't formally define effect sizes, let's not confuse the reader.
%
% ^ <yunfant@gmail.com> 2016-10-17T22:07:05.312Z.

The mean proportion of all OTUs across $G$ groups are equal if and only if $H_{0,A}$ is true for all $A \in \mathcal{I}$. Therefore, we define the global null as $H_0 = \cap_{A \in \mathcal{I}} H_{0,A}$. Controlling the Type-I error on the global null is simply equivalent to controlling the family-wise error rate (FWER) across $H_{0,A}$'s.

The following theorem makes controlling FWER straightforward: 
\begin{thm} \label{thm1}
Let $p_A$ be the MoM p-value for testing $H_{0,A}$. Under the global null $H_0 = \cap_{A \in \mathcal{I}} H_{0,A}$, $p_A$'s are asymptotically mutually independent as the number of subjects in each group goes to infinity.
\end{thm}

The appendix has a proof of this theorem.

The independence of p-value under the null grants one of the following procedures to control the exact FWER at level $\alpha$: (i) Sidak's procedure, in which one assigns equal Type I error $\alpha(A) = 1- (1-\alpha)^{1/\mathcal{I}}$ to all $A$'s (ii) allocate $\alpha_A$ according to the tree structure while constraining $1-\prod_{A \in \mathcal{I}}(1-\alpha_A) = \alpha$. After choosing the individual Type I error thresholds, one can report the collection of nodes $\{A: p_A < \alpha_A, A \in \mathcal{I}\}$ as being significant.

\section{PhyloScan: scan statistic over the tree tuples} \label{sec:Scan statistic over the tree tuples}
Cross-group difference in distributions of taxa counts often occurs in clusters or chains on the phylogenetic tree. If one internal node exhibits significant difference in relative proportion across several groups, then this is often associated with signals from at least one of its children or parent. Figure \ref{fig:PhyloDM triplets} shows four examples of signal clusters on American Gut data using the top 100 OTUs with the highest counts. In each graph, subjects are divided into two groups according to different ingestion frequencies in one of the following diets: milk and cheese, seafood, sugary sweets and vegetable. (details in Section \ref{sec:Application to American Gut dataset-Cross-group comparison}). The size of the circle on internal node $A$ is proportional to -$\log(p_A)$ from the cross-group comparison (the circle colors are irrelevant here). It is apparent that large circles tend to form in chaining patterns, which motivates scanning for signals in chains or clusters instead of on each node separately. Moreover the partitioning nature of the phylogenetic tree always leads to much smaller sample size on the bottom nodes (farthest from the root placed on top). Sharing information across nodes would alleviate the limitation to detect distributional differences on the bottom level.

Without prior knowledge of the length and shape of signal clusters, we only focus on triplets formulated by a certain internal node, its parent and one of its children. Each triplet has its own statistic defined as the sum of all the node statistics within, pooling signal strength from its members. The maximum of these statistics on all the triplets is then used to test the global null hypothesis. Our method belongs to the class of scan statistics [\cite{Glaz}], in which one searches for signals over varying sizes of windows. In our case, each window denotes a particular branch of the phylogenetic tree. The shape of our designed triplet reflects our knowledge of correlated signals on the tree, while the size of the triplet achieves a compromise between signal pooling around neighboring nodes and the ability to detect alternatives in short chains. Since the exact distribution of the maximum statistic is unknown, we design a novel method to calculate the upper and lower bound of its tail probability using low dimensional integrals that can be efficiently evaluated through standard numerical integration techniques. Since this entire hypothesis testing procedure is established on the phylogenetic tree decomposition, we call it PhyloScan. 
% * <marlee1982@gmail.com> 2016-10-15T15:18:53.398Z:
% 
% > Our method belongs to the class of scan statistics [\cite{Glaz}], in which one searches for the signals over varying sizes of windows.
%
% After this sentence, add the following sentence
% ``In our case the windows span over branches of the phylogenetic tree.''
%
% ^ <yunfant@gmail.com> 2016-10-17T22:34:13.815Z:
% 
% I added "In our case, each window denotes a particular branch of the phylogenetic tree" instead. Thought it might be easier to understand
% 
% ^ <marlee1982@gmail.com> 2016-10-18T17:36:04.042Z:
%
% Looks good now.
%
% ^ <yunfant@gmail.com> 2016-10-18T18:36:06.665Z.
% * <marlee1982@gmail.com> 2016-10-15T15:17:56.911Z:
%
% > its children
%
% --> ``one of its children''
%
% ^ <yunfant@gmail.com> 2016-10-18T16:06:58.320Z.
% * <marlee1982@gmail.com> 2016-10-15T15:14:43.913Z:
%
% > In this section, we devise another strategy that shares the same spirit of borrowing strength from nodes close to each other. 
%
% Remove this sentence. After removing the previous segment this sentence becomes redundant.
%
% ^ <yunfant@gmail.com> 2016-10-17T22:40:53.553Z.

\subsection{Overview} \label{sec:Scan statistic over the tree tuples-Overview}
For each $A \in \mathcal{I}$ such that $R(A) \in \mathcal{I}$ and $\mathcal{C}(A) \cap \mathcal{I} \neq \emptyset$, we define a triplet to be the set of three consecutive internal nodes $\{A, R(A),\mathcal{C}(A)_i \}$ where $i \in \{1,2,...,k(A)\}$ satisfies $\mathcal{C}(A)_i \in \mathcal{I}$. Let $\mathcal{B}$ be the set of all such triplets, and without loss of generality we write $\mathcal{B} = \{\mathcal{B}_1, \mathcal{B}_2, ..., \mathcal{B}_b\}$ where each $\mathcal{B}_i$ is a triplet and $b = |\mathcal{B}|$ depends on both $K$ and the structure of the tree. We assume the ordering of elements in $\mathcal{B}$ obeys the following rule: $\{A, R(A),\mathcal{C}(A)_i \}$ always has a smaller index than (or appear in front of) $\{\tilde{A}, R(\tilde{A}),\mathcal{C}(\tilde{A})_j \}$ if $\tilde{A} \subset A$. Now we proceed to define the test statistic for $\mathcal{B}_i$ as follows. First, each of the p-values on the internal nodes can be inverted to a chi-square random variable with $1$ degree of freedom, namely
$$Z_A = F^{-1}_1(p_A) \text{ for all } A \in \mathcal{I},$$
where $F_j$ denotes the cumulative distribution function (CDF) of $\chi^2_j$ distribution. Theorem \ref{thm1} states that under the global null $H_0$, $Z_A$'s are asymptotically mutually independent. In order to test the following hypothesis on each triplet 
$$H_{0, \mathcal{B}_i} = \bigcap\limits_{A \in \mathcal{B}_i} H_{0,A} \text{ vs } H_{a, \mathcal{B}_i}: \text{otherwise},$$
we define the statistic to be the sum of $Z_A$'s within:
\begin{equation} \label{Wi_from_ZA}
W_i = \sum\limits_{A \in \mathcal{B}_i} Z_A \text{ for } i = 1, 2, ..., b.
\end{equation}

It is apparent that each $W_i \sim \chi^2_3$ under $H_{0, \mathcal{B}_i}$. For the global null hypothesis $H_0 = \cap_{A \in \mathcal{I}}H_{0,A} = \cap_{i=1}^{b}H_{0,\mathcal{B}_i}$, we use the maximum of $W_i$'s as the test statistic:
\begin{equation} \label{W_from_Wi}
W = \max_{1\leq i \leq b} W_i.
\end{equation}

Since $\mathcal{B}_i$ overlaps with each other, $W_i$'s are heavily correlated and the exact distribution of $W$ is hard to derive. For  testing purposes, it suffices to calculate the tail probability of $W$. Suppose our observed value of the maximum statistic is $w$, and let $B_i(w) = \{W_i > w\}$ be the event of $i$th triplet statistic exceeding $w$ . Without incurring any confusion, we may drop $w$ and simply write $B_i$. We are mainly interested in the global p-value $P(\bigcup_{i=1}^{b} B_i)$, which boils down to the problem of bounding the union probability.

The simplest upper bound of the union probability is the Bonferroni inequality:
$$P(\bigcup_{i=1}^{b} B_i) \leq \sum\limits_{i=1}^{b} P(B_i)$$

There has been vast literature providing sharper bounds over the Bonferroni inequality in the past few decades. The results in \cite{Hunter}, \cite{Worsley} and \cite{Efron} suggest the following improvement:
\begin{align}
P(\bigcup_{i=1}^{b} B_i) &= P(B_1) + P(B_2 \cap B_1^c) + P(B_3 \cap B_1^c \cap B_2^c) + ... \nonumber \\
&\leq P(B_1) + \sum\limits_{i=2}^b \min_{j<i} P(B_i \cap B_j^c) \label{Bon1}
\end{align}
In particular, $\min_{j<i} P(B_i \cap B_j^c)$ is achieved at $j = i-1$ when the neighboring variables $(W_{i-1}, W_i)$ have the highest pairwise correlation. Each of the term inside summation can be easily evaluated by numerical integration. It can be easily generalized to the union of more than two sets to improve approximation. 
More generally, the above inequality belongs to the class of approximations with the following representation:
% * <yunfant@gmail.com> 2016-10-03T18:42:31.262Z:
%
% > More generally, 
%
% 123
%
% ^ <yunfant@gmail.com> 2016-10-03T18:43:00.353Z.
\begin{equation}\label{BonfGen}
P(\bigcup_{i=1}^b B_i) \leq \sum_{J \in S} (-1)^{|J|-1}f(J)P(\bigcap_{j \in J}B_j)
\end{equation}
where $f(J)$ is some non-negative function on subset of $S = \{1,2,...,b\}$. \cite{Naiman1} and \cite{Naiman2} gave results regarding when (\ref{BonfGen}) achieves equality. Following their work, \cite{Dohmen1} and \cite{Dohmen2} gave further improvement on the Bonferroni inequalities. There is also research on Bonferroni inequalities for particular applications, such as \cite{Dohmen and Tittmann} on partition lattice and \cite{Taylor} on maxima over Gaussian random fields. 

\subsection{Bounding the union probability} \label{sec:Scan statistic over the tree tuples-Bounding the union probability}
Our upper bound of the union probability involves a decomposition of $\bigcup_{i=1}^b B_i$ into (i) a union of independent events and (ii) their complement in $\bigcup_{i=1}^b B_i$. The probability of the union of independent events can be exactly evaluated, while a similar strategy to (\ref{Bon1}) is applied to estimate its complement. 

Specifically, let $\mathcal{M} = \{\mathcal{M}_1, \mathcal{M}_2, ..., \mathcal{M}_m\}$ be a class of disjoint nonempty subsets of $\mathcal{I}$ satisfying $\forall i \leq m, \exists j \leq b \text{ s.t } \mathcal{M}_i \subset \mathcal{B}_j$. For each $i$, define $M_i = \{ \sum_{A \in \mathcal{M}_i} Z_A > w \}$ to be the exceeding event on $\mathcal{M}_i$. It follows that $\forall i \leq m, \exists j \leq b \text{ s.t } M_i \subset B_j$. Write $M = \bigcup_{i=1}^{m}M_i$. This leads to
\begin{equation} \label{UP1}
P(\bigcup\limits_{i=1}^{b} B_i) = P(M) + P(M^c) \cdot P(\bigcup_{i=1}^{b} B_i | M^c )
\end{equation}
since $M \subset \bigcup_{i=1}^{b}B_i$. The independence of $M_i$'s leads to a straightforward calculation of $P(M)$ as $P(M) = 1 - F_1(w)^{t_1}F_2(w)^{t_2}F_3(w)^{t_3}$ where $t_l = |\{i \leq m: |\mathcal{M}_i|=l \}|$ and $F_i(\cdot)$ is the CDF of $\chi^2_i$ distribution. Next, we approximate $P(\bigcup_{i=1}^{b} B_i | M^c )$ using a similar strategy to (\ref{Bon1}). It is apparent that enlarging $M$ will always decrease $P(\bigcup_{i=1}^{b} B_i \cap M^c )$ and most likely the error of its upper bound, which makes our strategy superior to directly applying (\ref{Bon1}) to the $B_i$'s. 

The next question is how to choose an $M$ as large as possible. An obvious optimality condition is that $\bigcup_{i=1}^m \mathcal{M}_i = \mathcal{I}$, because otherwise we can always enlarge $M$ by $\{Z_A > w\}$ for a certain $A \in \mathcal{I} \setminus \bigcup_{i=1}^m \mathcal{M}_i $. Moreover, the elements in $\mathcal{M}$ should not be able to combine together and still belong to a certain element in $\mathcal{B}$, i.e. $\forall i_1,i_2 \leq m,\nexists j \leq b$ s.t. $\mathcal{M}_{i_1} \cup \mathcal{M}_{i_2} \subset \mathcal{B}_j$. This is because merging $\mathcal{M}_{i_1}$ and $\mathcal{M}_{i_2}$ enlarges $M$, i.e. $M_{i_1} \cup M_{i_2} \subset \{\sum_{A \in \mathcal{M}_{i_1} \cup \mathcal{M}_{i_2}} Z_A > w\}$. Since an exhaustive search over all combinations is computationally infeasible for large trees, we propose the following greedy algorithm that satisfies these optimality conditions:

\begin{enumerate}
\item[(a)] Order the elements in $\mathcal{I}$ as $A_{1}, A_{2}, ..., A_{|\mathcal{I}|}$ such that each internal node always appears in front of its children.
\item[(b)] Set $\mathcal{M} = \emptyset$ and $i =1$.  
\item[(c)] For each $i = 1, 2, ..., |\mathcal{I}|$, sequentially go through the following steps:
	\begin{enumerate}
	\item[(i)] If $\exists j \leq m $ s.t. $A_{i} \in \mathcal{M}_j$, set $i \leftarrow i + 1$ and go back to (c).
	\item[(ii)] If $\exists j_1, j_2 $ s.t. $\mathcal{C}(A_{i})_{j_1} \in \mathcal{I}$ and $\mathcal{C}\big( \mathcal{C}(A_{i})_{j_1}\big)_{j_2} \in \mathcal{I}$, set $\mathcal{M} \leftarrow \mathcal{M} \cup \{A_{i}, \mathcal{C}(A_{i})_{j_1},  \mathcal{C}\big(\mathcal{C}(A_{i})_{j_1} \big)_{j_2} \}$ and $i \leftarrow i + 1$. Go back to (c).
	\item[(iii)] If $\exists j_1$ s.t. $\mathcal{C}(A_{i})_{j_1} \in \mathcal{I}$, set $\mathcal{M} \leftarrow \mathcal{M} \cup \{A_{i}, \mathcal{C}(A_{i})_{j_1}\}$ and $i \leftarrow i + 1$. Go back to (c).
	\item[(iv)] Otherwise, set $\mathcal{M} \leftarrow \mathcal{M} \cup \{A_{i}\}$ and $i \leftarrow i + 1$. Go back to (c).
	\end{enumerate}
\end{enumerate}

The above greedy algorithm seeks to incorporate the longest chain (with maximum of 3 nodes) starting from $A_{i}$ and use its descendants as subsequent nodes, if  $A_{i}$ has not been included in $\mathcal{M}$ so far. Since the parent node is always considered ahead of its children, the resulting $\mathcal{M}$ will always satisfy the two aforementioned optimality conditions. As the algorithm prioritizes longer chains at each step, it effectively produces a large $M$ that yields relatively accurate estimates of the union probability for our applications (numerical results to be shown later).
% * <marlee1982@gmail.com> 2016-10-15T15:23:31.478Z:
%
% > result
%
% --> ``resulting''
%
% ^ <yunfant@gmail.com> 2016-10-17T22:42:26.521Z.

\begin{figure}[h]
\centering\includegraphics[height=7cm]{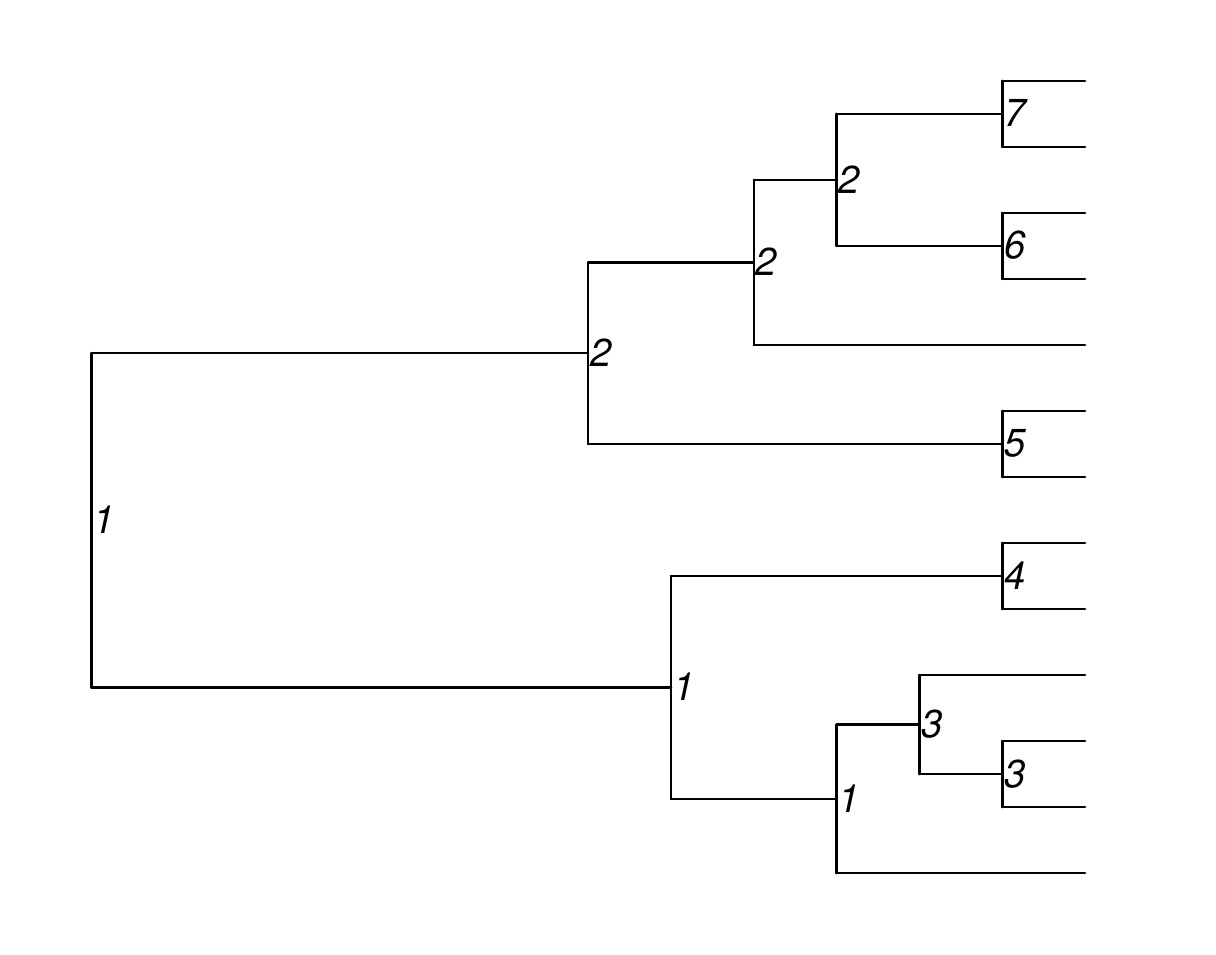}
\caption{\label{fig:example configuration} Example configuration of $\mathcal{M}$ using the greedy algorithm.}
\end{figure}

Figure \ref{fig:example configuration} shows an example of $\mathcal{M}$ on a simple phylogenetic tree with $K = 13$ OTUs. Each internal node in $\mathcal{M}_i$ is assigned the same number $i$ for $i = 1,2, ...,7$.

The remaining task is to put an upper bound on $P(\bigcup_{i=1}^{b} B_i | M^c )$. For each $i \leq b$, let $\mathcal{N}_i = \{j: |\mathcal{B}_j \cap \mathcal{B}_i| = 2 \text{ and } j < i\}$. Apparently $|\mathcal{N}_i| \leq 2 $ for all $i$ because of the ordering of $\mathcal{B}_i$'s. Write $B_{\mathcal{N}_i} = \bigcup_{j \in \mathcal{N}_i} B_j$ for short. Now we proceed as follows:
\begin{equation}\label{UP2}
P(\bigcup_{i=1}^{b} B_i | M^c ) \leq \sum\limits_{i=1}^b P(B_i \cap B_{\mathcal{N}_i}^c |M^c ) 
\end{equation}

The equation in (\ref{UP2}) is very similar to (\ref{Bon1}) in that for each $B_i$, it only includes the highest correlated events $B_{\mathcal{N}_i}$ in order to minimize the right side of the equation. It is worth noting that $P(B_i | M^c) = 0$ if $\mathcal{B}_i \in \mathcal{M}$, hence the strategy of prioritizing triplets while constructing $\mathcal{M}$. To efficiently evaluate each of the terms in the right side of (\ref{UP2}), notice that the distributions of $Z_A$ conditioned on $M^c$ are the same as the product of a truncated chi-square and an independent Dirichlet random variable, so their density function can be expressed using chi-square CDFs. Let $f_i(\cdot)$ and $F_i(\cdot)$ denote the density and CDF of $\chi^2_i$ distribution respectively, then the marginal density of $Z_A$ conditional on $M^c$ becomes
\begin{equation}\label{PDF1}
  f_A(z|M^c)=\begin{cases}
               \frac{f_1(z)}{F_1(w)}, \text{ if } |\mathcal{M}(A)| = 1 \wedge w \leq z\\
               \frac{F_1(w-z)f_1(z)}{F_2(w)}, \text{ if }|\mathcal{M}(A)| = 2 \wedge w \leq z\\
               \frac{F_2(w-z)f_1(z)}{F_3(w)}, \text{ if }|\mathcal{M}(A)| = 3 \wedge w \leq z\\
               0, \text{ otherwise}
            \end{cases}
\end{equation}
% * <marlee1982@gmail.com> 2016-10-15T17:20:45.175Z:
% 
% > \begin{equation}\label{PDF1}
% >   f_A(z|M^c)=\begin{cases}
% >                \frac{f_1(z)}{F_1(w)}, |\mathcal{M}(A)| = 1 \wedge w \leq z\\
% >                \frac{F_1(w-z)f_1(z)}{F_2(w)}, |\mathcal{M}(A)| = 2 \wedge w \leq z\\
% >                \frac{F_2(w-z)f_1(z)}{F_3(w)}, |\mathcal{M}(A)| = 3 \wedge w \leq z\\
% >                0, \text{ otherwise}
% >             \end{cases}
% > \end{equation}
% in the conditions, need to add ``if'' before ``$\mathcal{M}(A)|=$ '' etc.
% 
% ^ <yunfant@gmail.com> 2016-10-17T22:44:39.188Z.
where we define $\mathcal{M}(A) = \mathcal{M}_i$ if $A \in \mathcal{M}_i$. The existence and uniqueness of $\mathcal{M}(A)$ is guaranteed by the fact that $\{\mathcal{M}_1, \mathcal{M}_2, ..., \mathcal{M}_m\}$ are disjoint with $\bigcup_{i=1}^m \mathcal{M}_i = \mathcal{I}$. 

The joint density of $Z_{A_1}$ and $Z_{A_2}$ for $\forall A_1, A_2 \in \mathcal{I}$ is
\begin{equation}\label{PDF2}
  f_{A_1,A_2}(z_1,z_2|M^c)=\begin{cases}
               f_{A_1}(z_1|M^c)f_{A_2}(z_2|M^c), \text{ if }\mathcal{M}(A_1) \cap \mathcal{M}(A_2) = \emptyset\\
               \frac{\prod\limits_{i=1}^2 f_1(z_i)}{F_2(w)},\text{ if }\mathcal{M}(A_1) = \mathcal{M}(A_2) \wedge |\mathcal{M}(A_1)| = 2  \\
               \hspace{16mm} \wedge \sum\limits_{i=1}^2 z_i \leq w\\
               \frac{F_1(w-\sum\limits_{i=1}^2 z_i)\prod\limits_{i=1}^2 f_1(z_i)}{F_3(w)}, \text{ if } \mathcal{M}(A_1) = \mathcal{M}(A_2)  \\
               \hspace{34mm} \wedge  |\mathcal{M}(A_1)| = 3 \wedge \sum\limits_{i=1}^2 z_i \leq w \\
               0, \text{ otherwise}
               
            \end{cases}
\end{equation}
% * <marlee1982@gmail.com> 2016-10-15T17:22:35.596Z:
%
% > \begin{equation}\label{PDF2}
% >   f_{A_1,A_2}(z_1,z_2|M^c)=\begin{cases}
% >                f_{A_1}(z_1|M^c)f_{A_2}(z_2|M^c), \mathcal{M}(A_1) \cap \mathcal{M}(A_2) = \emptyset\\
% >                \frac{\prod\limits_{i=1}^2 f_1(z_i)}{F_2(w)},\mathcal{M}(A_1) = \mathcal{M}(A_2) \wedge |\mathcal{M}(A_1)| = 2 \wedge \sum\limits_{i=1}^2 z_i \leq w\\
% >                \frac{F_1(w-\sum\limits_{i=1}^2 z_i)\prod\limits_{i=1}^2 f_1(z_i)}{F_3(w)}, \mathcal{M}(A_1) = \mathcal{M}(A_2) \wedge  |\mathcal{M}(A_1)| = 3 \\
% >                \hspace{32mm} \wedge \sum\limits_{i=1}^2 z_i \leq w \\
% >                0, \text{ otherwise}
% >                
% >             \end{cases}
% > \end{equation}
%
% Same as the previous equation, need to at ``if'' in the first three conditions.
%
% ^ <yunfant@gmail.com> 2016-10-17T22:47:02.646Z.

Given $w$, we pre-calculate the density functions in (\ref{PDF1}) and (\ref{PDF2}), and the CDFs of $Z_A$ using (\ref{PDF1}) and of $Z_{A_1} + Z_{A_2}$ using $(\ref{PDF2})$ up to a certain precision and store them into the memory. This turns each term in the right side of (\ref{UP2}) into at most two-dimensional integrals. We evaluate these integrals using the functions \verb|cuhre|  and  \verb|suave| in R package \verb|R2Cuba| [\cite{Hahn}].

Plugging (\ref{UP2}) into (\ref{UP1}) gives
\begin{equation}\label{UP3}
P_0 = P(\bigcup_{i=1}^{b} B_i ) \leq P_U = P(M) + P(M^c) \cdot \sum_{i=1}^b P(B_i \cap B_{\mathcal{N}_i}^c |M^c ),
\end{equation}
where $P_0$ is the actual p-value and $P_U$ is its upper bound. Let $\epsilon_U = P_U - P_0$ be the error of our approximation. Using a similar strategy to Theorem A1 in \cite{Taylor}, it follows that
\begin{align}
\epsilon_U &=P(M^c) \sum_{i=1}^b \big( P(B_i \cap B_{\mathcal{N}_i}^c |M^c ) - P(B_i \cap B_{i-1}^c \cap B_{i-2}^c \cap ... B_{1}^c |M^c )  \big) \nonumber \\
&= P(M^c) \sum_{i=1}^b P(B_i \cap B_{\mathcal{N}_i}^c \cap \big(\bigcup_{j<i, j\notin \mathcal{N}_i}B_j \big)  |M^c ) \nonumber \\
&\leq P(M^c) \sum_{i=1}^b P(\bigcup_{j<i, j\notin \mathcal{N}_i} ( B_i \cap B_j )  |M^c ) \nonumber \\
&\leq P(M^c) \sum_{i=1}^b\sum_{j<i, j\notin \mathcal{N}_i} P(B_i \cap B_j  |M^c ) \label{UP4}
\end{align}
Each term in (\ref{UP4}) can be evaluated by at most three dimensional numerical integral using the pre-calculated densities and CDFs. This also establishes 
\begin{equation}
P_0 \in \big(P_U - P(M^c) \sum_{i=1}^b\sum_{j<i, j\notin \mathcal{N}_i} P(B_i \cap B_j  |M^c ), P_U \big)  \label{UP5}
\end{equation}

In the next section we give the numerical results of $P_U$ and upper bound of $\epsilon_U$ using the phylogenetic tree from the American Gut dataset. In addition, we have the following theorem on the convergence rate of the relative error with regards to the observed statistic $w$.

\begin{thm} \label{thm2}
Given the set of all triplets $\mathcal{B}$ and the partition $\mathcal{M}$ on the internal nodes $\mathcal{I}$, define the following quantities:
\begin{itemize}
\item $\xi_1 = |\{(i,j): \mathcal{B}_i \cap \mathcal{B}_j = \emptyset, \mathcal{B}_i \notin \mathcal{M}, \mathcal{B}_j \notin \mathcal{M}\text{ and } 1\leq j<i\leq b |$
\item $\xi_2 = |\{(i,j): |\mathcal{B}_i \cap \mathcal{B}_j| = 1 , \mathcal{B}_i \notin \mathcal{M}, \mathcal{B}_j \notin \mathcal{M}\text{ and } 1\leq j<i\leq b |$
\item $\xi_3 = |\{i: |\mathcal{M}_i| = 3 \text{ and } 1\leq i\leq m \}| $
\end{itemize}

Then under the condition that $(\xi_3-1)\big(1-F_3(w_T)\big) <0.1$ and $w_T \geq 12$, we have
\begin{align*}
\frac{\epsilon_U}{P_U} &< \frac{\xi_1}{0.95\xi_3 + \xi_T}\big(1-F_3(w)\big) + \frac{0.9\xi_2}{0.95\xi_3+\xi_T}\sqrt{\frac{\pi}{2}}\cdot\frac{1}{w} \\
&= \mathcal{O}(e^{-\frac{w}{6}}) + \mathcal{O}(w^{-1}) 
\end{align*}
for all $w \geq w_T$, where 
$$\xi_T = \frac{\sum_{i=1}^b P(B_i \cap B_{\mathcal{N}_i}^c \cap M^c )}{1-F_3(w)} \text{ is evaluated at } w = w_T$$ 
\end{thm}

See the appendix for the proof.

\subsection{Comparison with Monte-Carlo simulation} \label{sec:Scan statistic over the tree tuples-Comparison with Monte-Carlo simulation}
We compared the lower and upper bound in (\ref{UP5}) with Monte-Carlo simulated p-values. Each round of simulation produces $5\times10^4$ simulated maximum triplet statistics, and we use their proportion of exceeding $w$ as the estimated p-value. The maximum triplet statistic is simulated through generating the $\chi^2_1$ distributed $Z_A$'s for all $A \in \mathcal{I}$ and then applying (\ref{Wi_from_ZA}) and (\ref{W_from_Wi}). We draw the comparison for a variety of scenarios with different numbers of OTU $K \in \{50,100\}$ and different observed statistic $w \in \{15,20,25\}$. Given $K$, the tree structure is obtained from keeping the top $K$ OTUs with the highest count in all feces samples from the American Gut dataset (introduced later). In each scenario, we provide the histogram of simulated p-values over 5000 rounds.

\begin{figure}[t!]
\centering
\subfloat{\includegraphics[height=4.15cm]{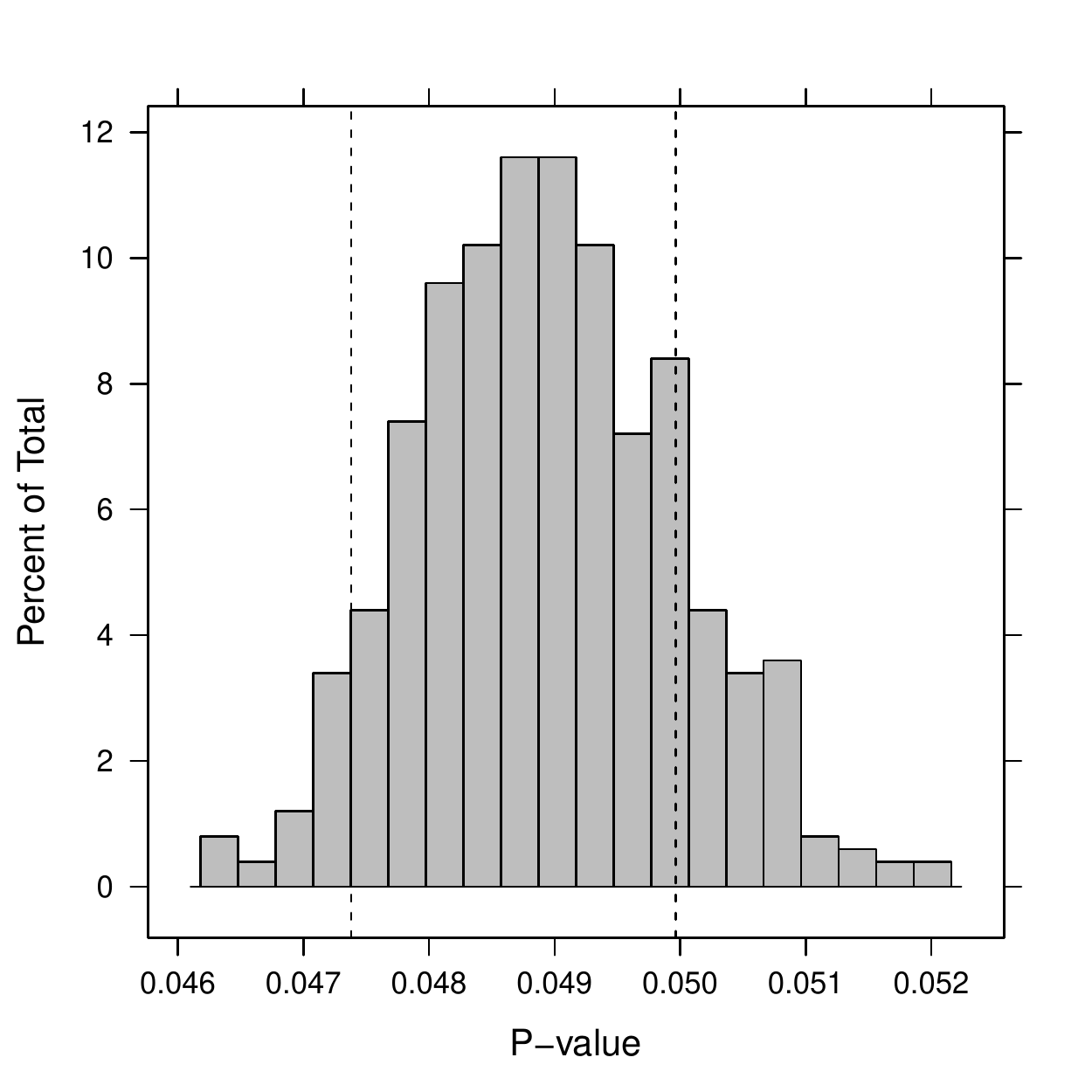}} 
\subfloat{\includegraphics[height=4.15cm]{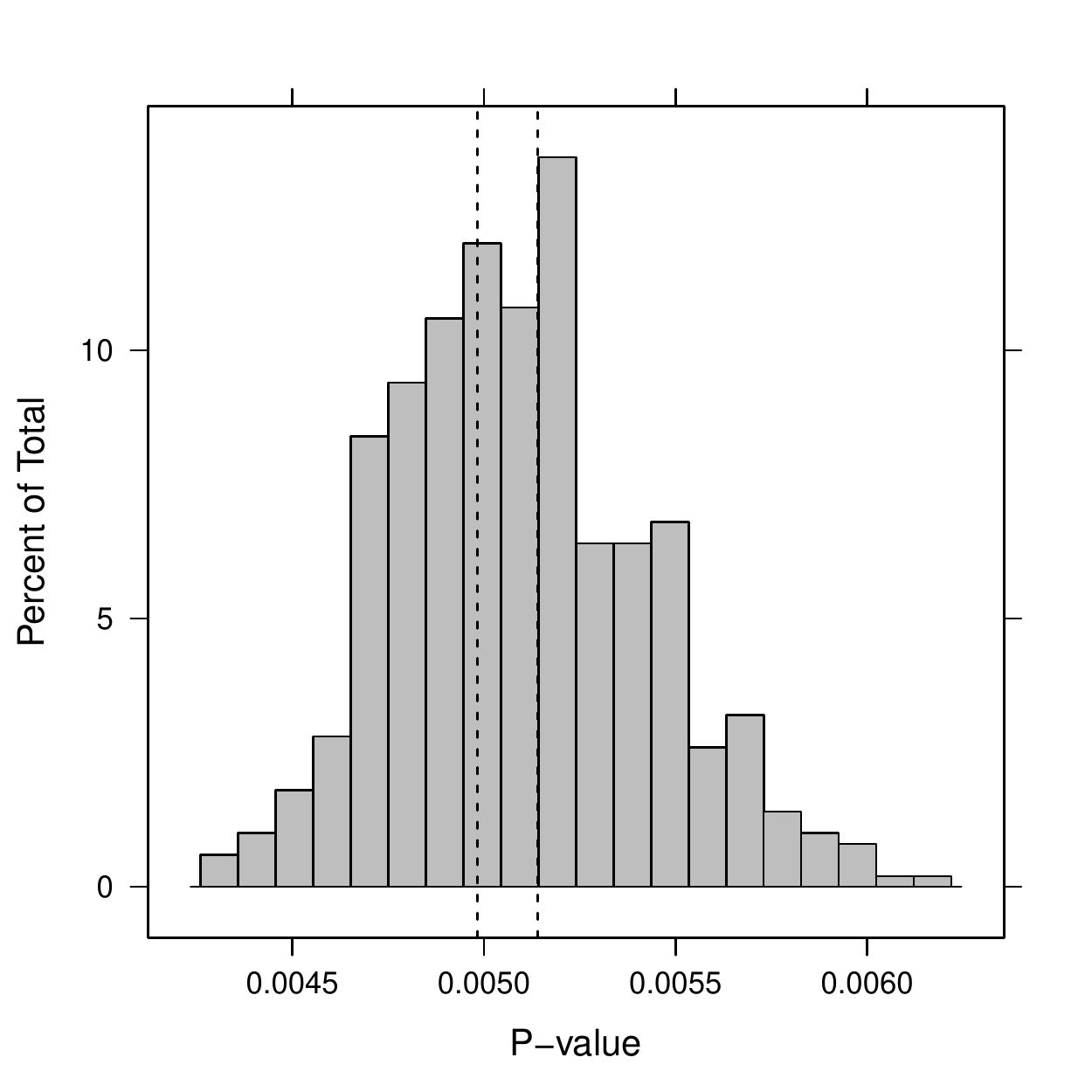}}
\subfloat{\includegraphics[height=4.15cm]{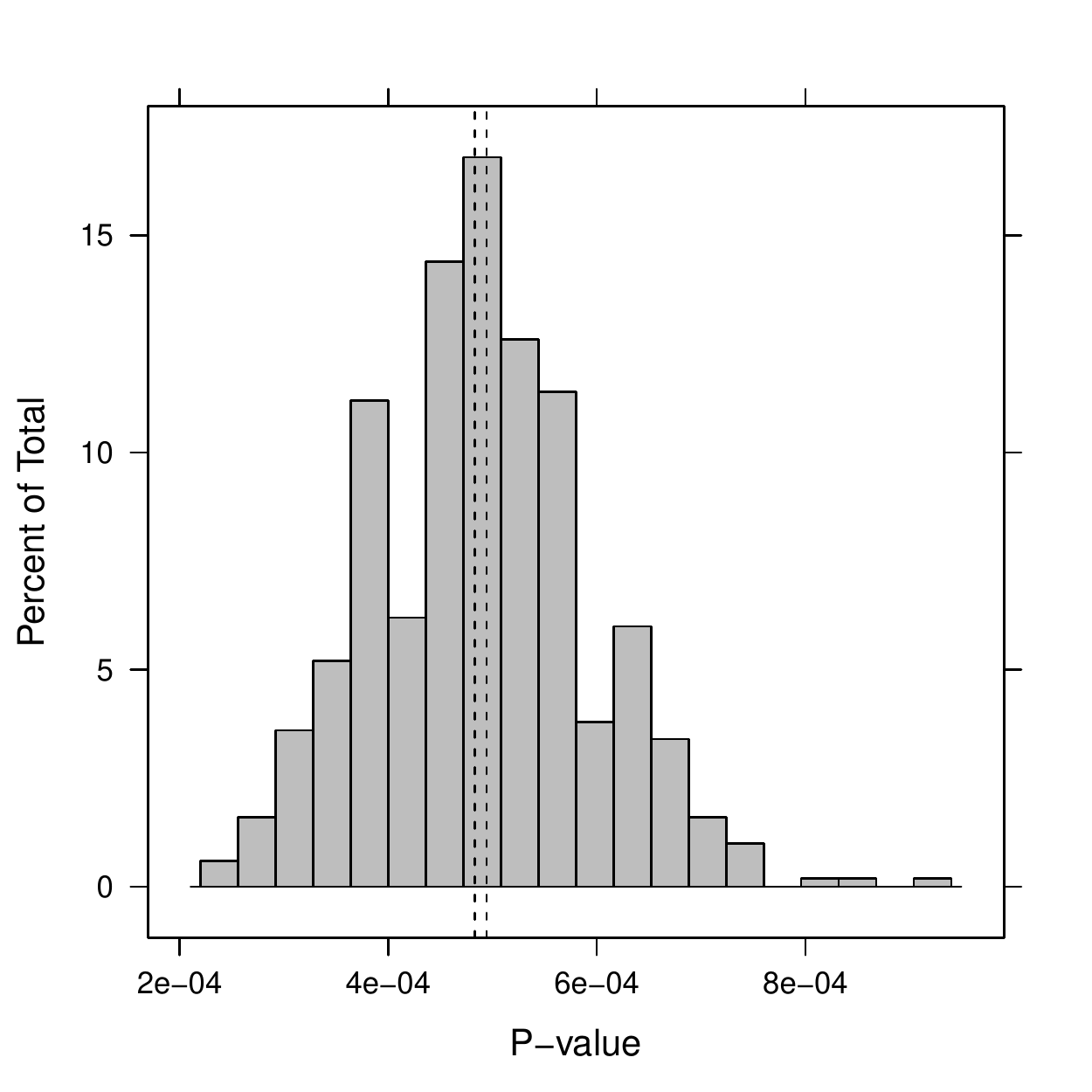}} \\ 
\subfloat{\includegraphics[height=4.15cm]{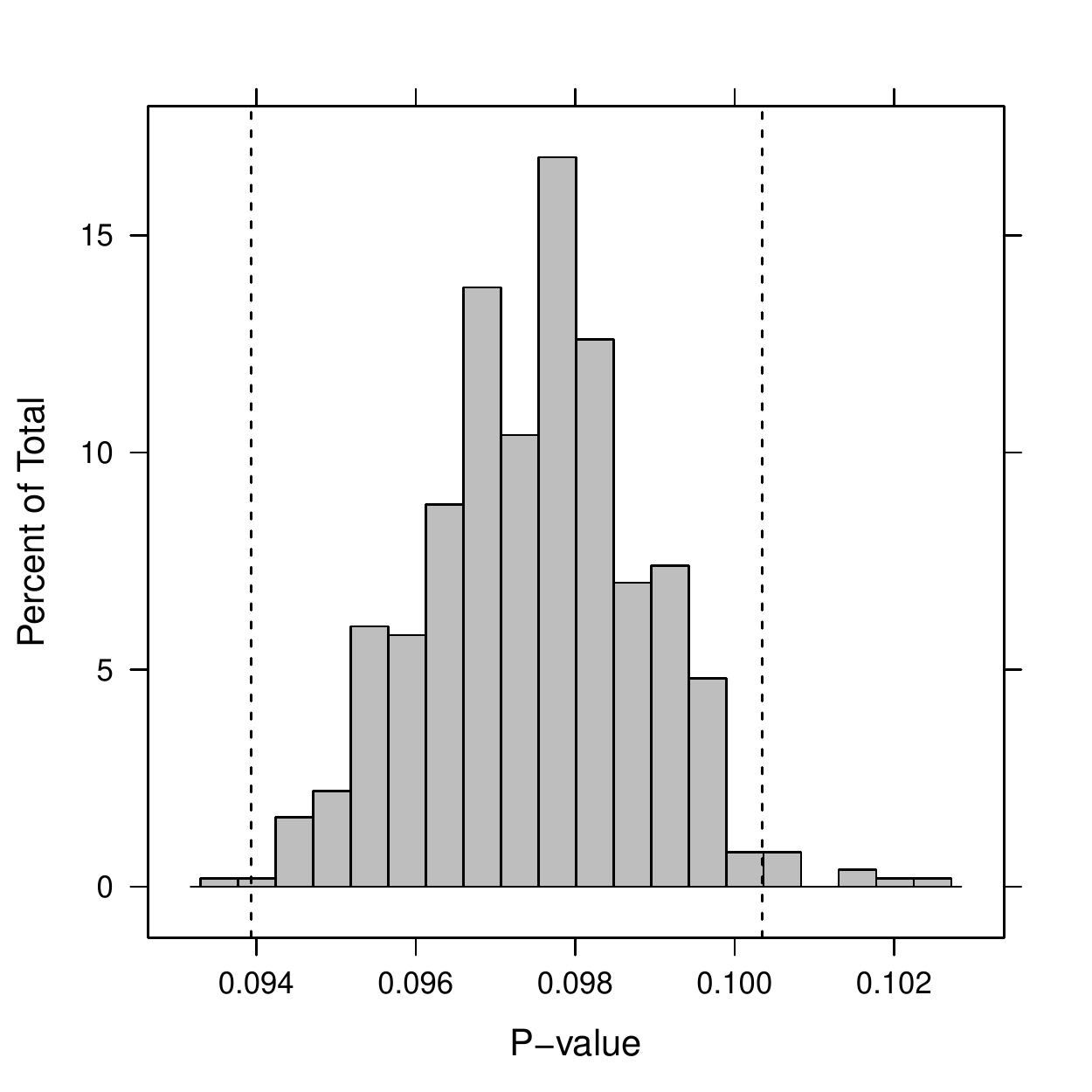}} 
\subfloat{\includegraphics[height=4.15cm]{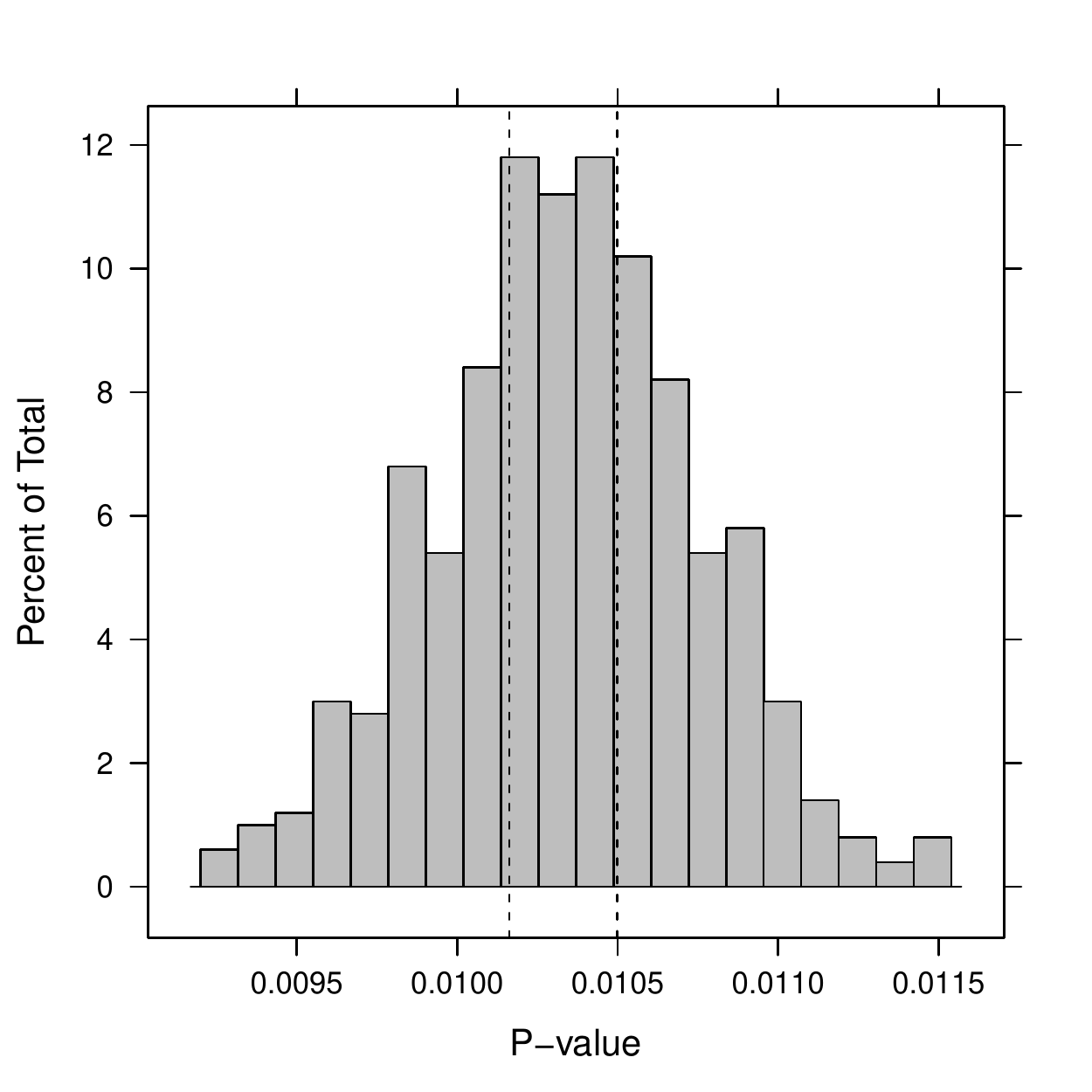}} 
\subfloat{\includegraphics[height=4.15cm]{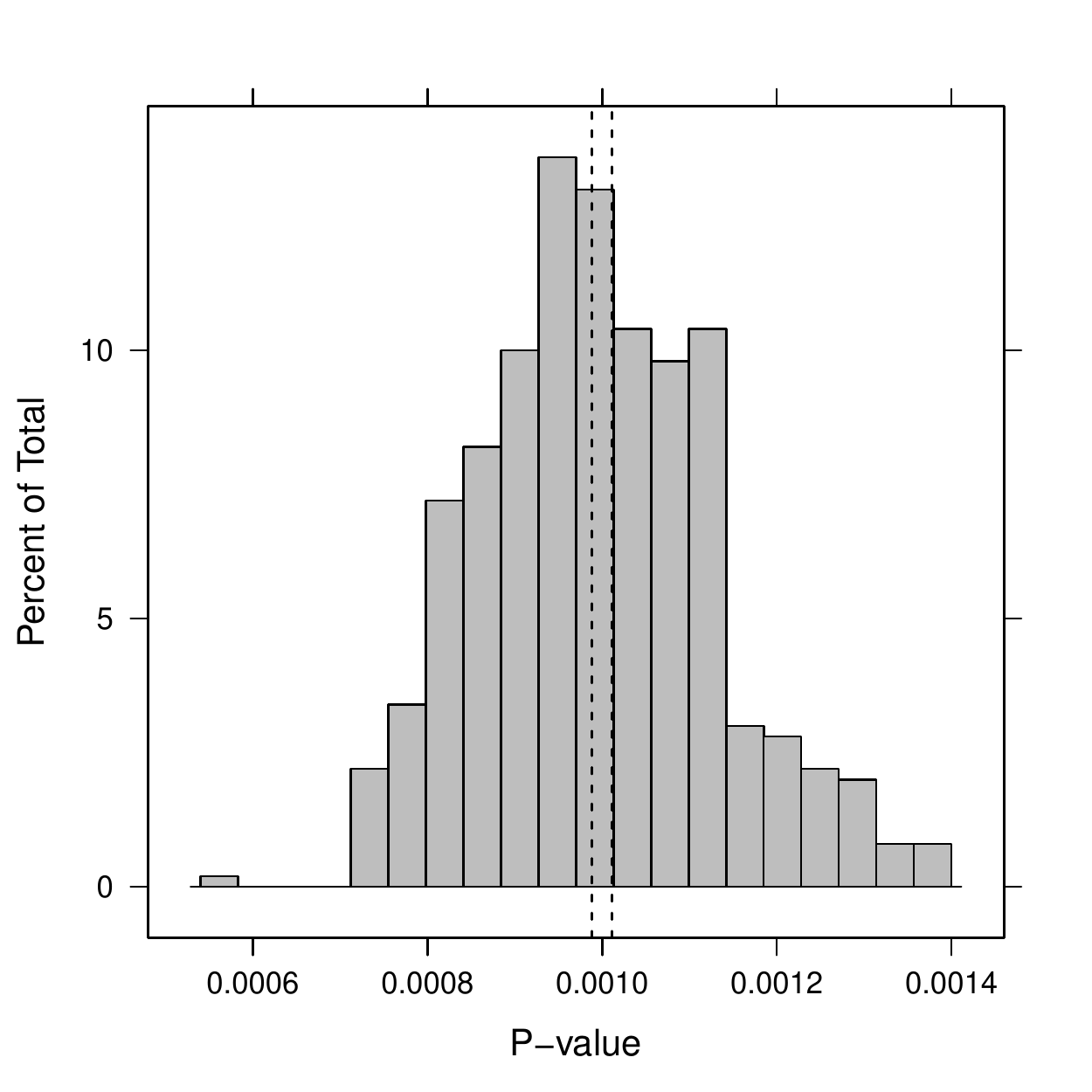}} \\
\caption{\label{fig:MC verification} Comparison between the interval bound and simulated p-values. Each simulated p-value is the proportion exceeding $w$ over $5\times10^4$ runs. Dashed lines indicate the upper and lower bound as in (\ref{UP5}). Top row and bottom row indicate $K = 50$ and $K = 100$ respectively. Left column: $w = 15$, middle column: $w = 20$, and right column: $w=25$.}
\end{figure}

Figure \ref{fig:MC verification} shows that our bounds consistently contain the center of the simulated p-values. Since the Monte Carlo p-values are merely binomial proportions, the ratio of their spread (measured by standard deviation) to $P_0$ goes to infinity as $P_0 \rightarrow 0$. In contrast, our method gives a ratio that tends to zero by Theorem \ref{thm2}. This makes our approach particularly useful for scenarios where a large number of tests leads to very small p-value threshold after multiple testing correction. In order to keep a fixed relative error, the computation time of Monte Carlo method needs to scale up much faster with $w$ than our method.

% * <marlee1982@gmail.com> 2016-10-15T15:26:52.468Z:
%
% > large number
%
% --> ``a large number''
%
% ^ <yunfant@gmail.com> 2016-10-17T22:48:17.850Z.
% * <marlee1982@gmail.com> 2016-10-03T17:48:23.361Z:
%
% > the
%
% remove
%
% ^ <yunfant@gmail.com> 2016-10-03T21:35:46.542Z.

\section{Application to American Gut Project} \label{sec:Application to American Gut dataset}
American Gut Project [\cite{McDonald}] is an open-access and crowd-sourced initiative that involves the public into the research of human microbiome and aims at providing a much more comprehensive reference set than the previous Human Microbiome Project [\cite{HMP}]. After contributing to the project fund, participants complete a questionnaire and ship their microbiome sample to the sequencing lab currently located at University of California, San Diego. The questionnaire covers a wide range of topics regarding demographic information, diet, lifestyle, etc. Sampling sites include skin, tongue and feces, although the vast majority of participants provided the feces sample. The samples are sequenced on 16s rRNA and further processed by QIIME [\cite{Caporaso}] pipeline to produce the OTUs and the phylogenetic tree. The 2016 May 16 cohort of public dataset includes more than eight thousands of subjects, with median of sequences per individual as 14680 and standard deviation as 32455.

\subsection{Cross-group comparison} \label{sec:Application to American Gut dataset-Cross-group comparison}
Our focus is comparison of the feces microbiome across different diet habits. We pick the top 100 OTUs with the highest count summing over all feces samples. The phylogenetic tree on these OTUs is fully binary. We also select a total of seven categories of diet from the questionnaire. Each diet divides the samples into two groups; group 1 consists of individuals with ingestion rate less than three times per week, and group 2 corresponds to more than or equal to three times per week. Since the questions are not compulsory, a large number of subjects do not leave any response. The diet names and their sample sizes in both groups are as follows: fermented plant (880 vs 3024), fruit (2336 vs 1660), milk and cheese (1743 vs 2261), poultry (1421 vs 2611), seafood (556 vs 3452), sugary sweet (1542 vs 2493) and vegetable (3422 vs 577).

For each diet type, we test the equality of mean proportions between two groups using three methods: DTM with PhyloScan, DTM with Sidak correction and global DM. Table \ref{table:AG all} presents their p-values using the 100 OTUs. The DTM(PhyloScan) column contains $P_U$ and the upper bound of its error $\epsilon_U$ in the parenthesis, both derived in Section \ref{sec:Scan statistic over the tree tuples-Bounding the union probability}. DTM(Sidak) column is calculated as the Sidak multiple testing correction $1-(1-\min_{A \in \mathcal{I}} p_A)^{|\mathcal{I}|} \approx |\mathcal{I}| \min_{A \in \mathcal{I}} p_A$. We also provide DM p-values after grouping the 100 OTUs into family and class levels, respectively. The grouping operation based on taxonomy is a common practice in recent papers including \cite{Rosa} and \cite{Chen}. At each taxonomic level, all OTUs with missing taxa information are placed into the same group. This leads to a total of 22 categories on family level and 9 categories on class level. The DM p-values are calculated using the R package \verb|HMP|. 

\renewcommand{\arraystretch}{1.2}
\begin{table}[h]
\footnotesize
\caption{\label{table:AG all} DM and DTM p-values for testing microbiome compositions across different diet habits. DTM(PhyloScan) contains $P_U$ and the upper bound on $\epsilon_U$ shown in parenthesis. DTM(Sidak) contains the Sidak-corrected p-values $1-(1-\min_{A \in \mathcal{I}} p_A)^{|\mathcal{I}|}$. For DM, we provide p-values directly on the 100 OTUs as well as after grouping the 100 OTUs into family and class levels, respectively.}
\centering
\begin{tabular}{c c c c c c c}
\hline 
&  \multicolumn{2}{c}{\textbf{DTM}} & & \multicolumn{3}{c}{\textbf{\Gape[0.2cm][0cm]{DM}}}\\ \cline{2-3} \cline{5-7}
\textbf{\Gape[0cm][0.2cm]{Diet}} & \textbf{PhyloScan}  & \textbf{Sidak} & & \textbf{OTU} &  \textbf{Family} & \textbf{Class} \\
\hline
\Gape[0.2cm][0cm]{Fermented plant} & 0.308 & 0.239 & & 0.377 & 0.147 & 0.038 \\ 
 & (0.036) \\
\Gape[0.2cm][0cm]{Fruit} & $8.75 \times 10^{-5}$ & $1.64 \times 10^{-4}$ & & $2.81\times 10^{-3}$ & 0.012 & 0.218\\
& $(1.52\times10^{-6})$ \\
\Gape[0.2cm][0cm]{Milk and cheese} & $1.07 \times 10^{-4}$ & $6.48 \times 10^{-3}$ & & 0.029 & 0.262 & 0.285 \\ 
 & $(1.86\times10^{-6})$ \\
\Gape[0.2cm][0cm]{Poultry} & 0.023 & 0.111 & & 0.158 & 0.287 & 0.691 \\
& $(8.71 \times 10^{-4})$ \\
\Gape[0.2cm][0cm]{Seafood} & $6.85\times10^{-5}$  & $6.40 \times 10^{-3}$ & & $1.75\times10^{-4}$ & 0.194 & 0.772  \\
& $(1.17\times10^{-6})$  \\
\Gape[0.2cm][0cm]{Sugary sweet} & $5.13 \times 10^{-3}$ & 0.015 & & 0.719 & 0.558 & 0.815 \\
& $(1.43\times10^{-4})$ \\
\Gape[0.2cm][0cm]{Vegetable} & $7.39 \times 10^{-5}$ & $4.79\times 10^{-5}$ & & $3.77 \times 10^{-3}$ & $1.88 \times 10^{-3}$ & 0.014\\
& \Gape[0cm][0.2cm]{$(1.27\times10^{-6})$} \\

\hline

\end{tabular}
\end{table}
\renewcommand{\arraystretch}{1}

All diet habit comparisons exhibit significant DTM(PhyloScan) p-values at 0.05 level except fermented plant. This is consistent with the findings in \cite{Turnbaugh} and \cite{David}, both of which established that the human gut microbiome is highly sensitive to the dietary nutrient composition. DTM(Sidak) also produces similar significance results. although in five out of seven comparisons its p-values are larger than PhyloScan. The largest relative difference occurs at seafood comparison (Sidak p-value about $100$ times greater than PhyloScan). The rest two comparisons (fermented plant and vegetable) are likely to have either a single dominating signal or weak clustering pattern, both of which hurt testing power after signal pooling. Still, PhyloScan only has mildly larger p-values under these circumstances. This data analysis concludes that PhyloScan is superior to Sidak correction in most cases. Note that p-values of DM on OTUs fail to reach significance for fermented plant, poultry and sugary sweet. This happens in even more comparisons on family and class levels.  

% * <marlee1982@gmail.com> 2016-10-15T15:29:01.040Z:
%
% > except
%
% --> ``except for''
%
% --> ``except for''
%
% ^ <yunfant@gmail.com> 2016-10-17T22:53:06.881Z:
% 
% I think both are fine here
% 
% ^ <yunfant@gmail.com> 2016-10-18T18:36:25.071Z.

\begin{figure}[t!]
\centering
\subfloat{\includegraphics[height=6.7cm]{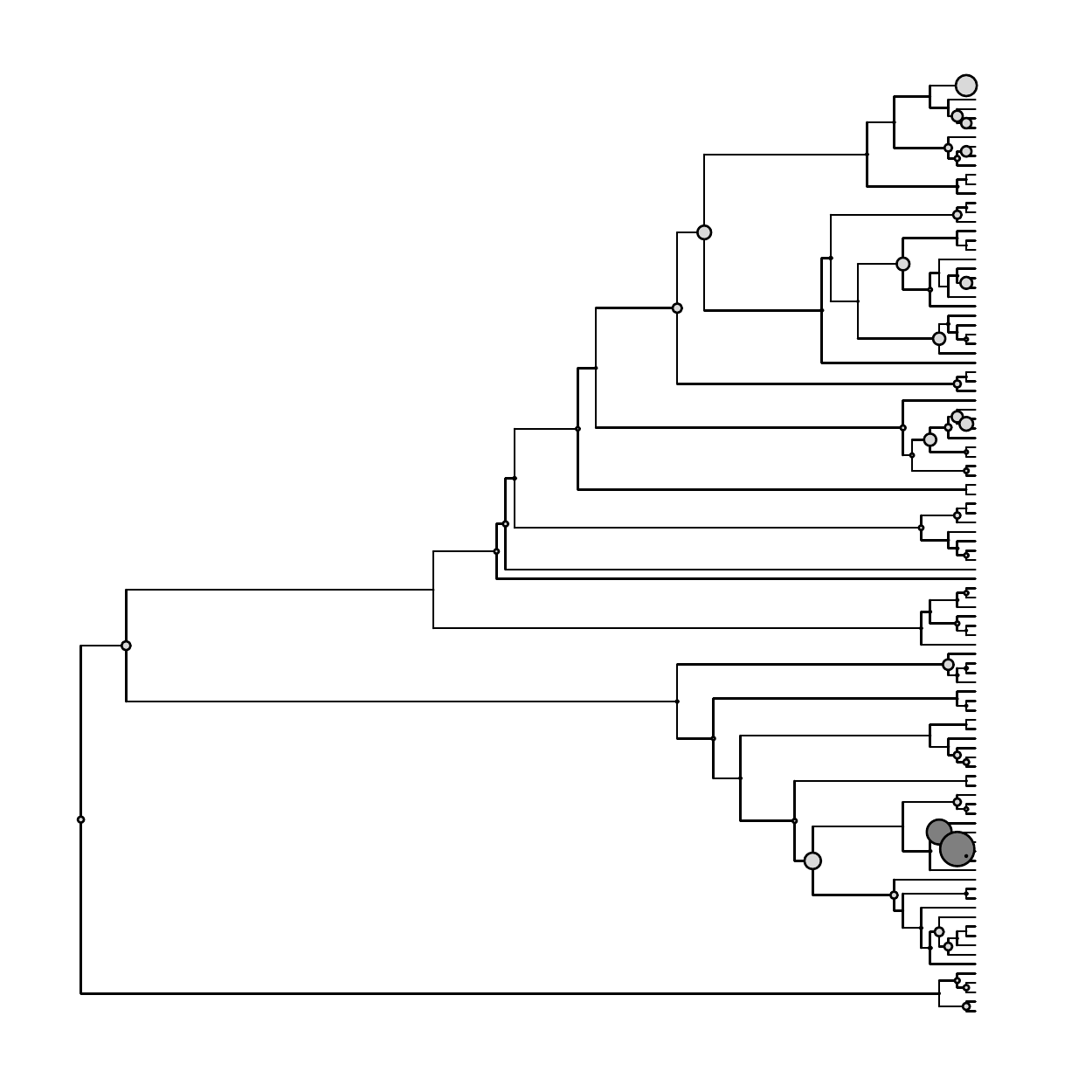}} 
\subfloat{\includegraphics[height=6.7cm]{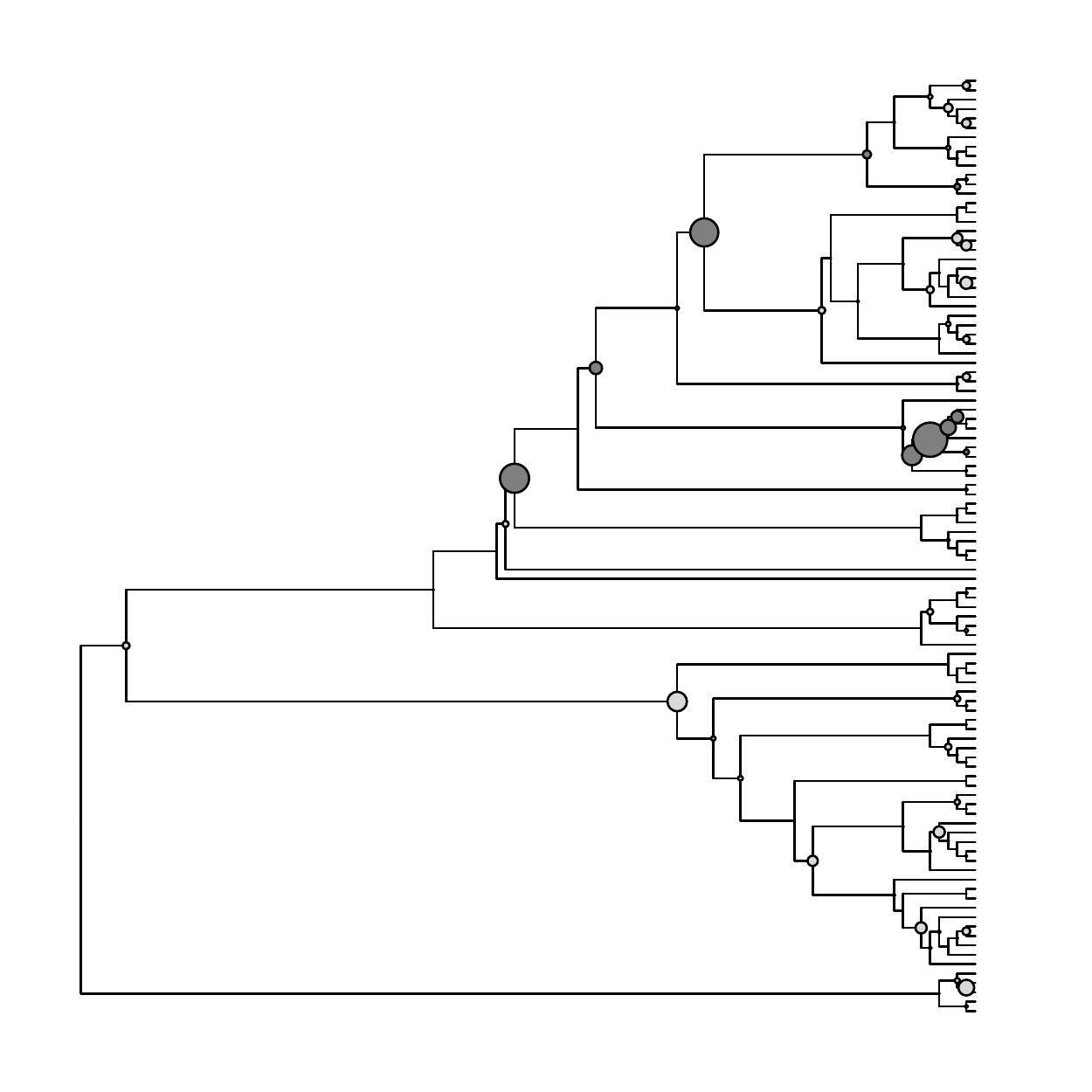}}  \\
\subfloat{\includegraphics[height=6.7cm]{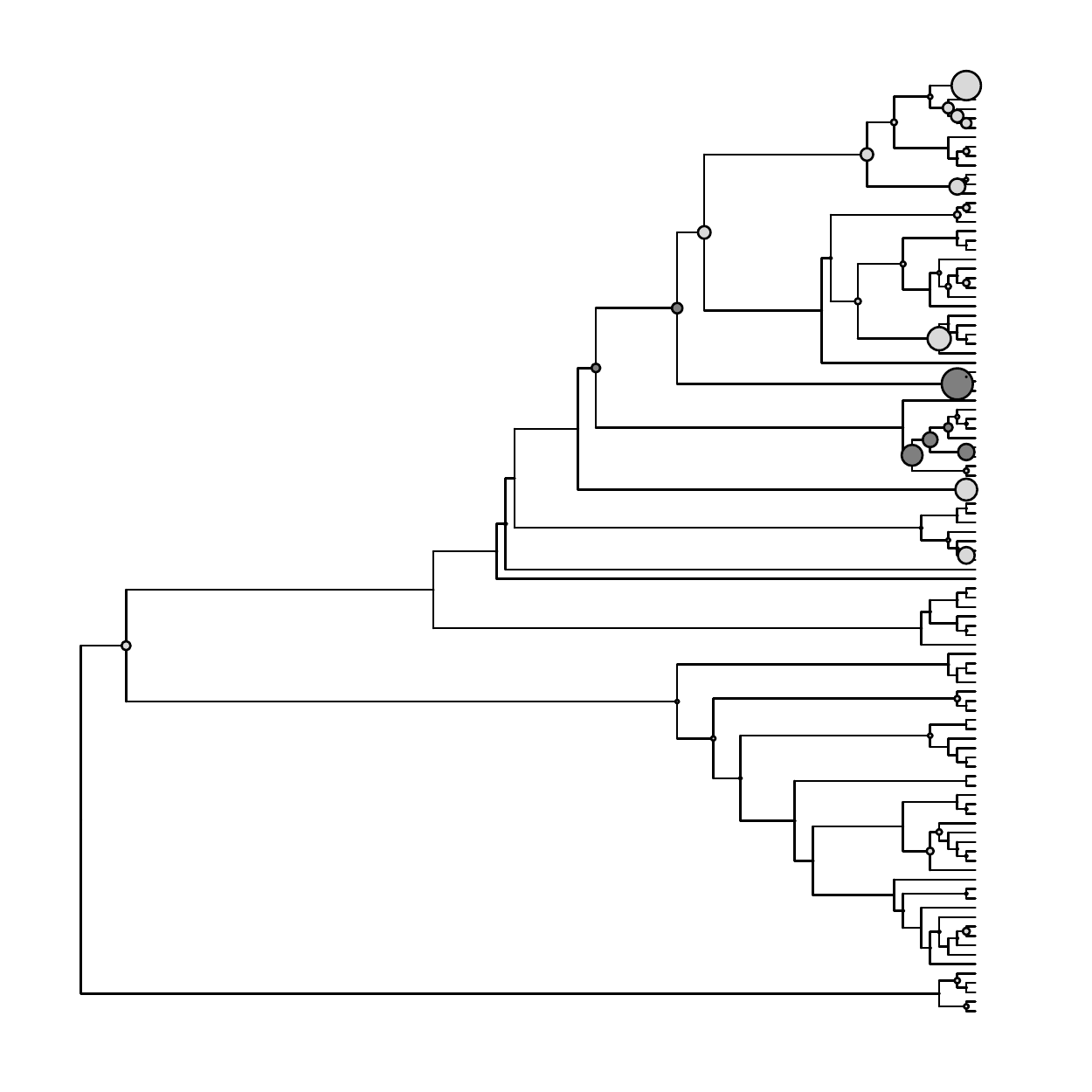}} 
\subfloat{\includegraphics[height=6.7cm]{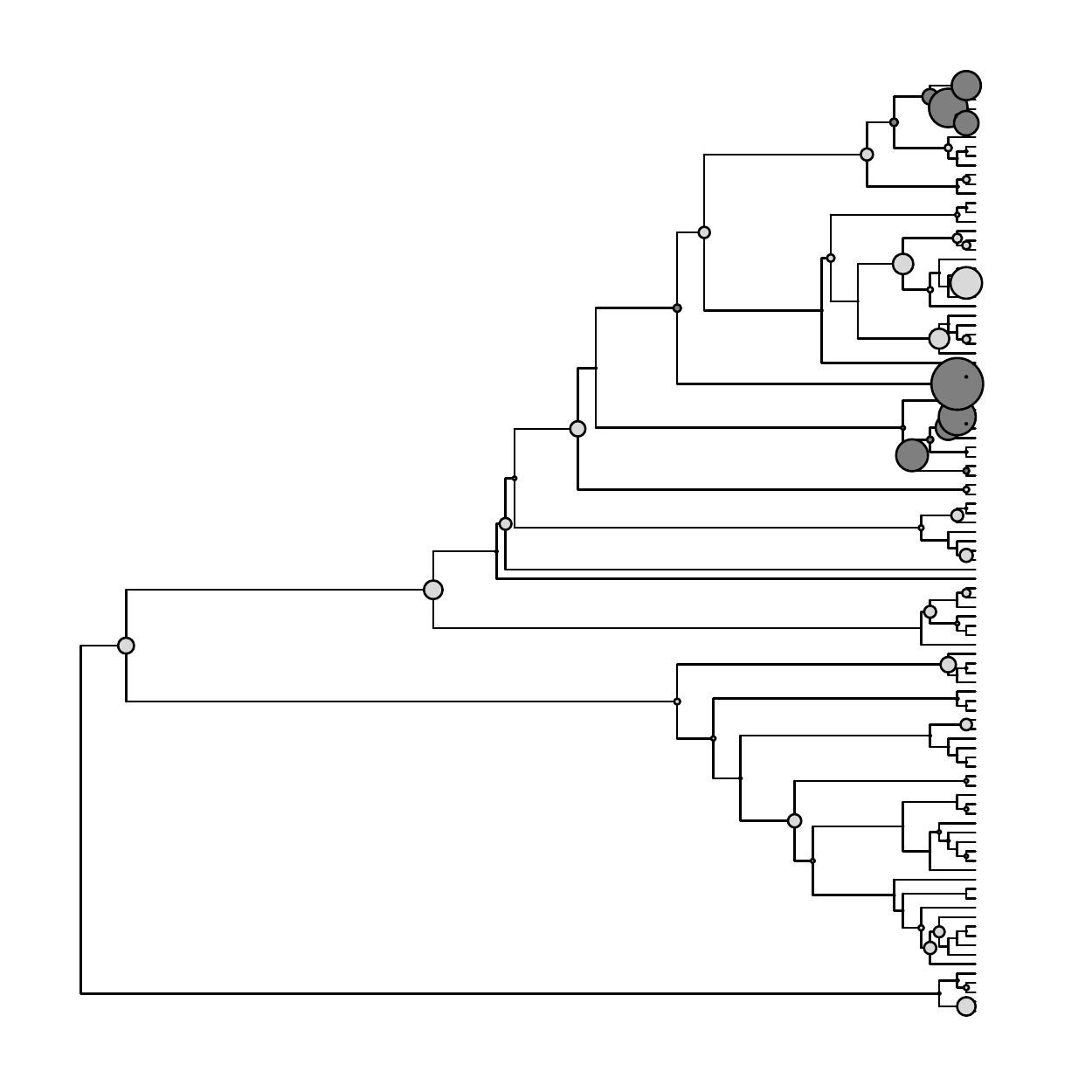}} 
\caption{\label{fig:PhyloDM triplets} Significant triplets from DTM testing. Top left: Milk and cheese, top right: seafood, bottom left: sugary sweets and bottom right: vegetable. The size of the circle on internal node $A$ is proportional to $-\log(p_A)$. Triplets with $W_i > 16.579$ are plotted in dark gray.}
\end{figure}

We further visualize the significant internal nodes in Figure \ref{fig:PhyloDM triplets} for four of the diet comparisons: milk and cheese, seafood, sugary sweets and vegetable. Using a simple binary search, we find that $w =16.579 $ yields $P_U = 0.05$ with $\epsilon_U \leq 2.43 \times 10^{-3} $. All triplets with test statistics greater than the above threshold, i.e. \textbf{$W_i > 16.579$}, are plotted in dark gray. In some cases the triplets overlap with each other, leading to a much longer chain than the original setup. We also provide the taxonomy for all internal nodes that belong to a certain significant triplet in Table \ref{table:PhyloDM taxa}. The internal node taxon is defined according to the following algorithm: starting from kingdom, repeatedly decrease the rank by one level until the descendant OTUs of that particular internal node no longer share the same taxa on the next lower rank (missing taxa on OTUs are excluded). The algorithm then picks the common taxon of the descendant OTUs on the rank at which the algorithm stops. In other words, the internal node taxon reflects the finest classification upon which all of its descendant OTUs agree. 

% * <marlee1982@gmail.com> 2016-10-15T15:30:25.902Z:
%
% > test statistic
%
% --> ``test statistics''
%
% ^ <yunfant@gmail.com> 2016-10-17T22:56:51.822Z.

\renewcommand{\arraystretch}{1.2}
\begin{table}[h]
\caption{\label{table:PhyloDM taxa} Taxa on significant triplets from PhyloDM hypothesis testing for each diet comparison. Each internal node that belongs to a certain significant triplet is assigned a taxon based on its descendant OTUs (details described in section \ref{sec:Application to American Gut dataset-Cross-group comparison}). Only the lowest level taxon is reported for each internal node. We omit the class rank since there are no significant internal nodes on such level in any cross-group comparisons.}
\centering
\footnotesize
\begin{tabular}{ c c c c c c c}
\hline
\textbf{Diet} & \Gape[0.2cm][0.2cm]{\textbf{Phylum}} & \textbf{Order} & \textbf{Family} & \textbf{Genus}  \\
\hline
\Gape[0.3cm][0cm]{Fermented plant} & --- & --- & --- & ---\\ 
\Gape[0.3cm][0cm]{Fruit} & Firmicutes & Clostridiales & Clostridiaceae & Clostridium \\
& --- & --- & Ruminococcaceae & Faecalibacterium\\
\Gape[0.3cm][0cm]{Milk and cheese} & --- & ---& ---& Bacteroides \\
\Gape[0.3cm][0cm]{Poultry} & --- & Clostridiales & --- & Coprococcus\\
\Gape[0.3cm][0cm]{Seafood} & Firmicutes & Clostridiales & Lachnospiraceae & Coprococcus \\
& --- & --- & Ruminococcaceae & Ruminococcus \\
\Gape[0.3cm][0cm]{Sugary sweet} & Firmicutes & Clostridiales & Lachnospiraceae &  Coprococcus \\
& --- & --- & Ruminococcaceae & ---  \\
\Gape[0.3cm][0cm]{Vegetable} & Firmicutes & Clostridiales & Lachnospiraceae & Blautia \\
& --- & --- & Ruminococcaceae & Coprococcus \\
& --- & --- & --- & \Gape[0cm][0.3cm]{Lachnospira} \\
\hline
\end{tabular}
\end{table}
\renewcommand{\arraystretch}{1}

%\multirow{3}{*}{\begin{tabular}[x]{@{}c@{}}Lachnospiraceae \\ Ruminococcaceae \end{tabular}}
%\begin{tabular}[x]{@{}c@{}}Foo\\bar\end{tabular}
%\multirow{3}{*}{\pbox{20cm}{Lachnospiraceae \\ Ruminococcaceae}}

\subsection{DM vs DTM test} \label{sec:Application to American Gut dataset-DM vs PhyloDM test}
We can also test the model fit of DTM against DM directly on the OTUs. Since DM is nested in the DTM family, we can use the likelihood ratio test (LRT) for 
$$H_0: \exists \nu>0 \text{ s.t. } \forall A \in \mathcal{I}, \nu_A = \nu\sum\limits_{\omega\in A} \pi_{\omega} \text{ (DM)} \text{ vs } H_a: \text{otherwise (DTM)}$$
with the test statistic defined as
\begin{equation} \label{DMvsPhyloDM}
\Lambda(\boldsymbol{x}) = -2\log\frac{\mathcal{L}(\hat{\nu}, \hat{\boldsymbol{\pi}})} {\mathcal{L}_T(\{(\hat{\nu}_A, \hat{\boldsymbol{\pi}}_A): A \in \mathcal{I} \})}  \sim \chi^2_{|\mathcal{I}|-1} \text{ under } H_0,
\end{equation}
where $(\hat{\nu}, \hat{\boldsymbol{\pi}})$ in the numerator of (\ref{DMvsPhyloDM}) are MLEs of the DM model, and each $(\hat{\nu}_A, \hat{\boldsymbol{\pi}}_A)$ in the denominator are obtained through maximizing the DTM conditional likelihood (\ref{PhyloDM1}). We use the low-storage BFGS optimization implemented in package \verb|nloptr| to calculate the MLE estimates. The degrees of freedom in (\ref{DMvsPhyloDM}) is $|\mathcal{I}|-1$ for a binary phylogenetic tree since (i) $\dim(\boldsymbol{\pi}) = \dim(\{ \boldsymbol{\pi_A}: A \in \mathcal{I} \}) = K - 1$, and (ii) $\dim(\{ \nu_A: A \in \mathcal{I} \}) = |\mathcal{I}|$. 

Table \ref{table:LRT} shows the LRT result. The test is separately applied to male and female Caucasians living in a variety of geographic regions. Each region is consisted of certain states in the U.S. defined according to Bureau of Economic Analysis. The degree of freedom for all tests is 98 since our phylogenetic tree is binary and $|\mathcal{I}| = |K| - 1 = 99$. All scenarios yield LRT p-values less than $10^{-10}$, which indicates significantly improved fit on the data using DTM. We also note that $\Lambda(x)$ in general increases with the sample size, as evidence towards heterogeneity in OTU dispersion strengthens with more available data.
% * <marlee1982@gmail.com> 2016-10-03T17:54:31.977Z:
% 
% > which indicates drastically improved fit on the data using PhyloDM
% A p-value only indicates statistical significance. it doesn't measure how much the fit is improved practically. when large enough sample size, even a small improvement in the fit in a practically sense can be highly significant statistically.
% 
% ^ <yunfant@gmail.com> 2016-10-04T17:50:42.729Z:
%
% Changed "drastically" to "significantly"
%
% ^ <yunfant@gmail.com> 2016-10-17T22:57:11.078Z.

\renewcommand{\arraystretch}{1.2}
\begin{table}[h]
\caption{\label{table:LRT} Likelihood ratio test for DM vs DTM. The test is separately applied to male and female Caucasians in a variety of geographic regions. Each test is accompanied by the LRT statistic $\Lambda(x)$ and the sample size. }
\footnotesize
\centering
\begin{tabular}{ c c c c c c}
\hline
 & \multicolumn{2}{c}{\Gape[0.2cm][0.2cm]{\textbf{Male}}} & &  \multicolumn{2}{c}{\textbf{Female}}\\ \cline{2-3} \cline{5-6}
\textbf{Region} & $\mathbf{\Lambda(\boldsymbol{x})}$ & \textbf{\Gape[0.2cm][0.2cm]{Sample size}} & & $\mathbf{\Lambda(\boldsymbol{x})}$ & \textbf{Sample size} \\
\hline
Far West & 14179.76 & 663  & & 15768.10 & 775 \\
Great Lakes & 4025.36 & 180 & & 5541.45 & 276\\
Mideast & 7497.80 & 328 & & 8112.80 & 396 \\
New England & 5630.21 & 239 & &  5793.38 & 269 \\
Rocky Mountain & 6084.90 & 244 & & 6676.25 & 300 \\
Southeast & 7030.43 & 324 & & 7383.72 & 366 \\
Southwest & 3442.69 & 153 & & 3675.52 & 189 \\
\hline
\end{tabular}
\end{table}
\renewcommand{\arraystretch}{1}

\subsection{Simulation} \label{sec:Application to American Gut dataset-Simulation} 
We use two simulation strategies to evaluate the power of PhyloScan test under various conditions. From  American Gut dataset, we extracted a total of 662 individuals who identified themselves as male Caucasian living in the far west (Alaska, California, Hawaii, Nevada, Oregon and Washington). In each round of simulation, these selected samples are randomly divided into two equal-sized groups to generate data under the global null. For data under the alternative, the first simulation strategy randomly selects an OTU and increases its count by a fixed percentage for all samples in the second group, whereas the second simulation strategy random selects an internal node and increases the count all of its descendant OTUs equally by a fixed percentage for all samples in the second group. We use the same 100 OTUs as before and produce 5000 rounds of simulation.
 % * <marlee1982@gmail.com> 2016-10-15T15:33:41.904Z:
%
% > increase
%
% --> ``increases'' this typo appeared twice
%
% ^ <yunfant@gmail.com> 2016-10-17T22:57:49.824Z.
% * <marlee1982@gmail.com> 2016-10-15T15:32:56.041Z:
%
% > increase
%
% --> ``increases''
%
% ^ <yunfant@gmail.com> 2016-10-17T22:57:50.921Z.

\begin{figure}[h]
\centering
\subfloat{\includegraphics[height=5cm]{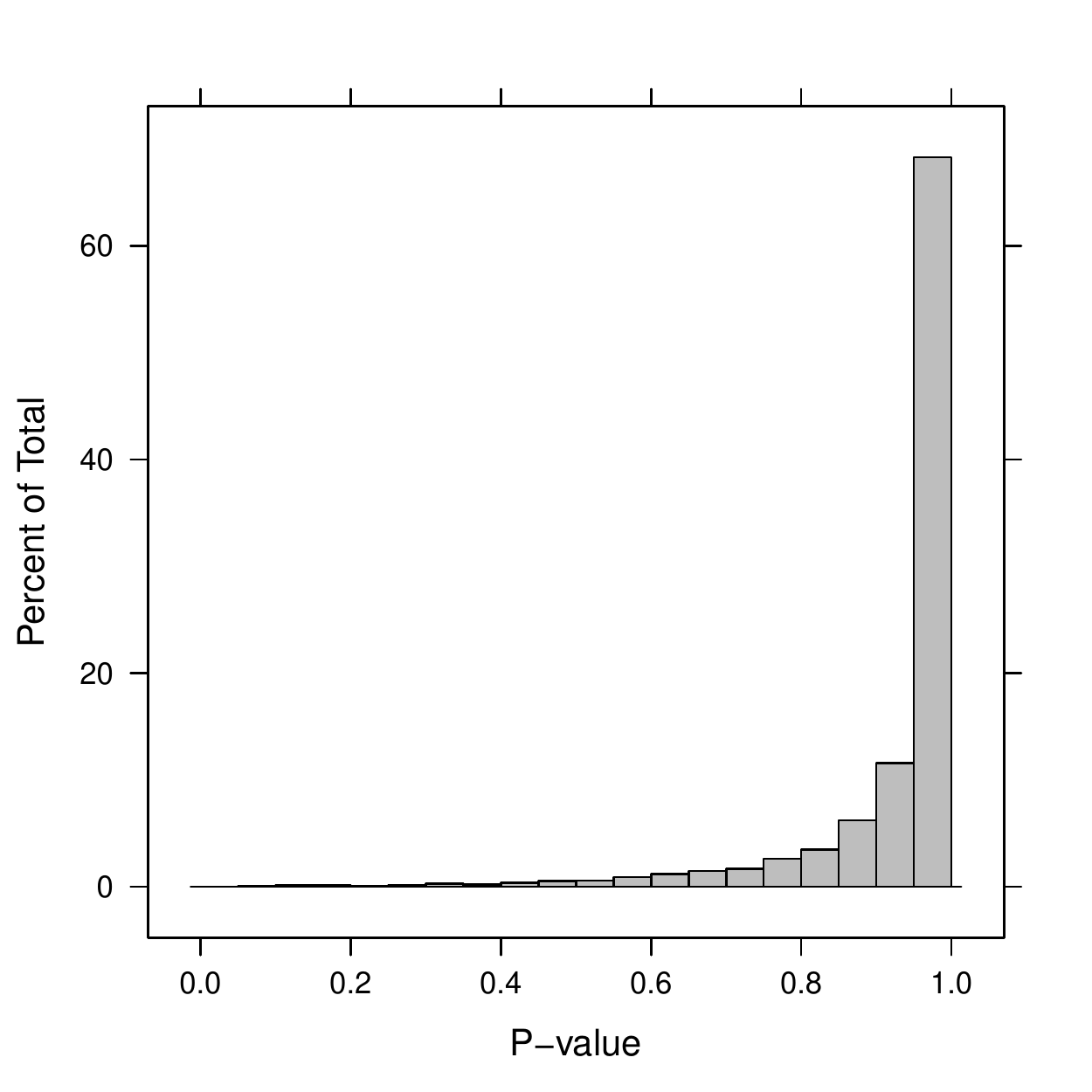}} 
\subfloat{\includegraphics[height=5cm]{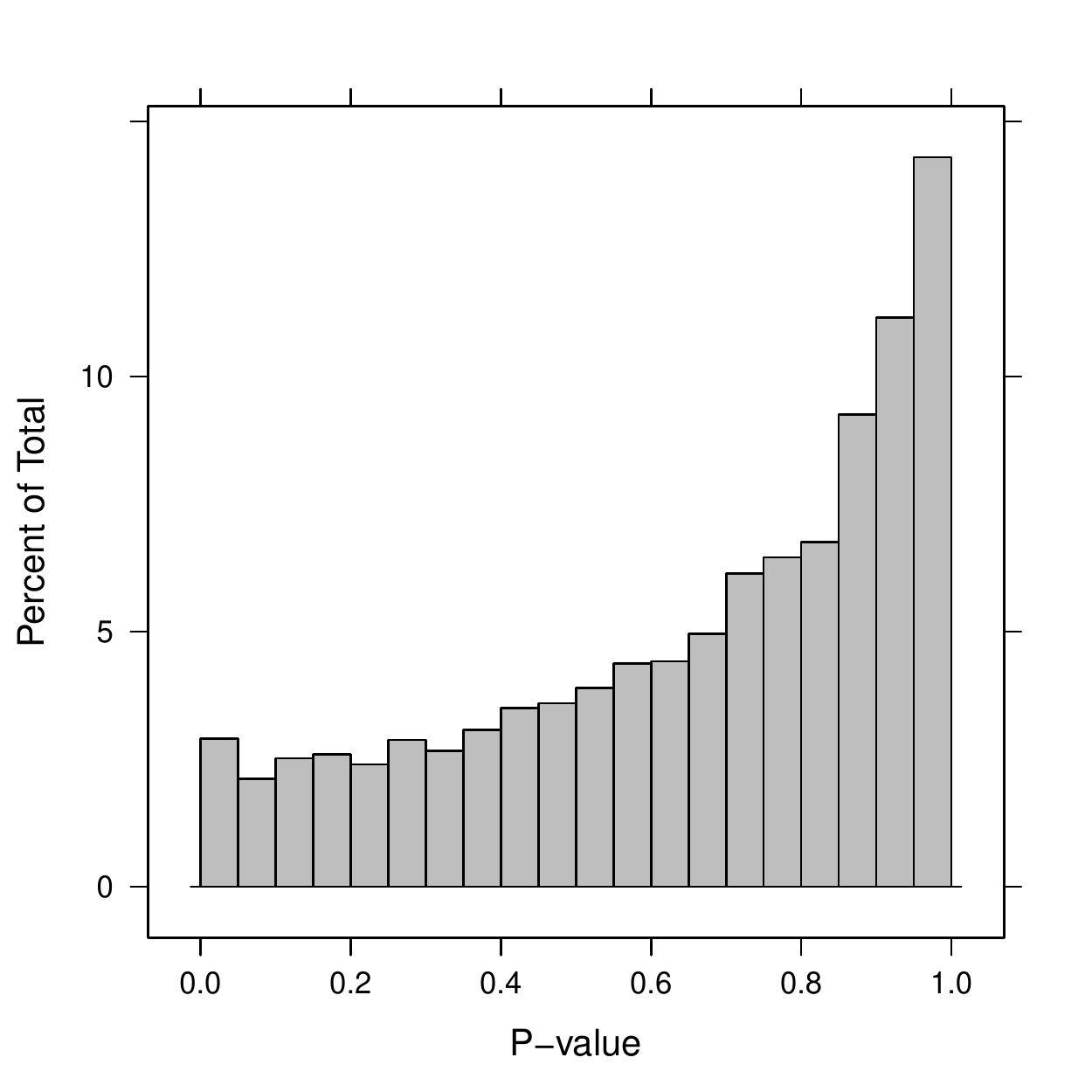}}  \\
\subfloat{\includegraphics[height=5cm]{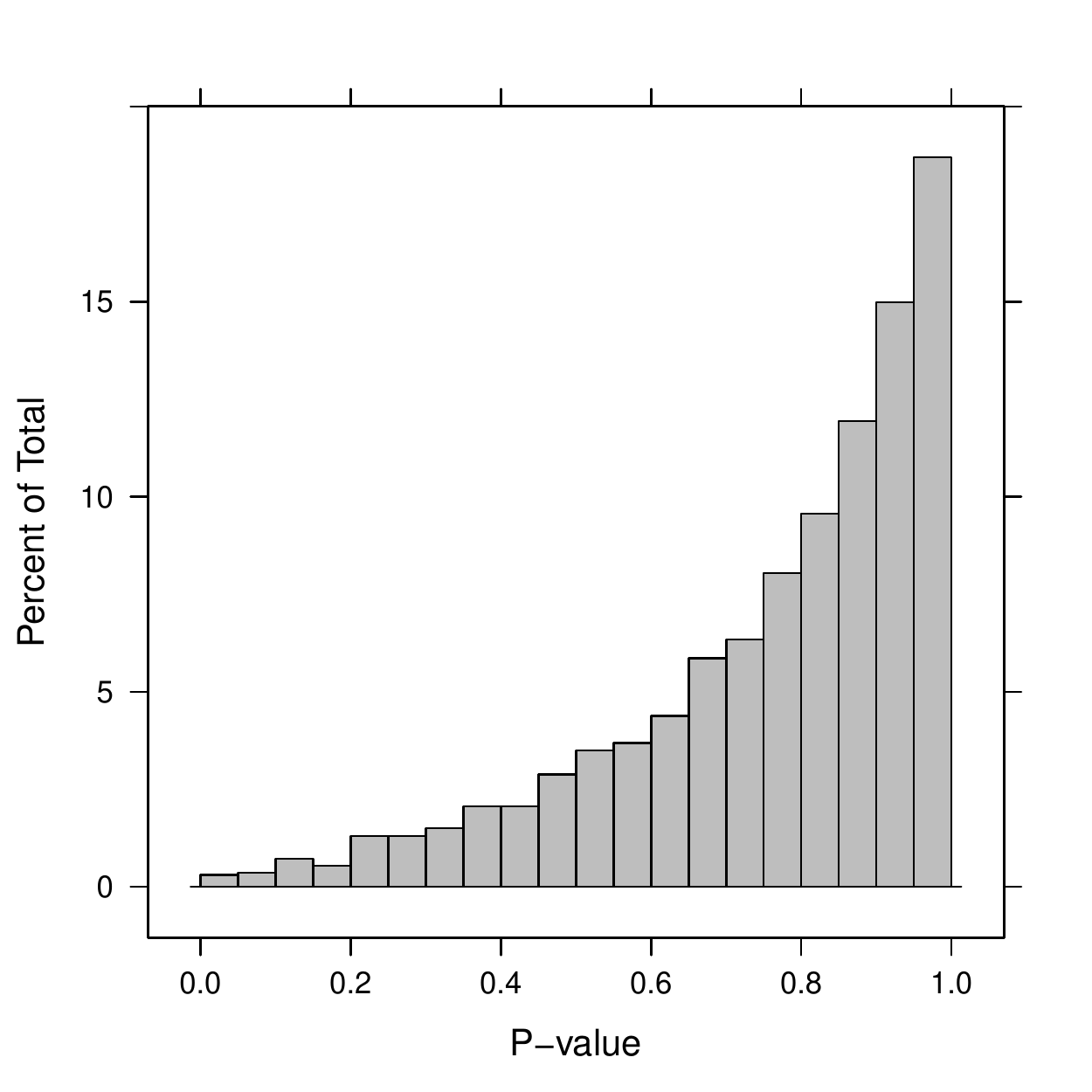}} 
\subfloat{\includegraphics[height=5cm]{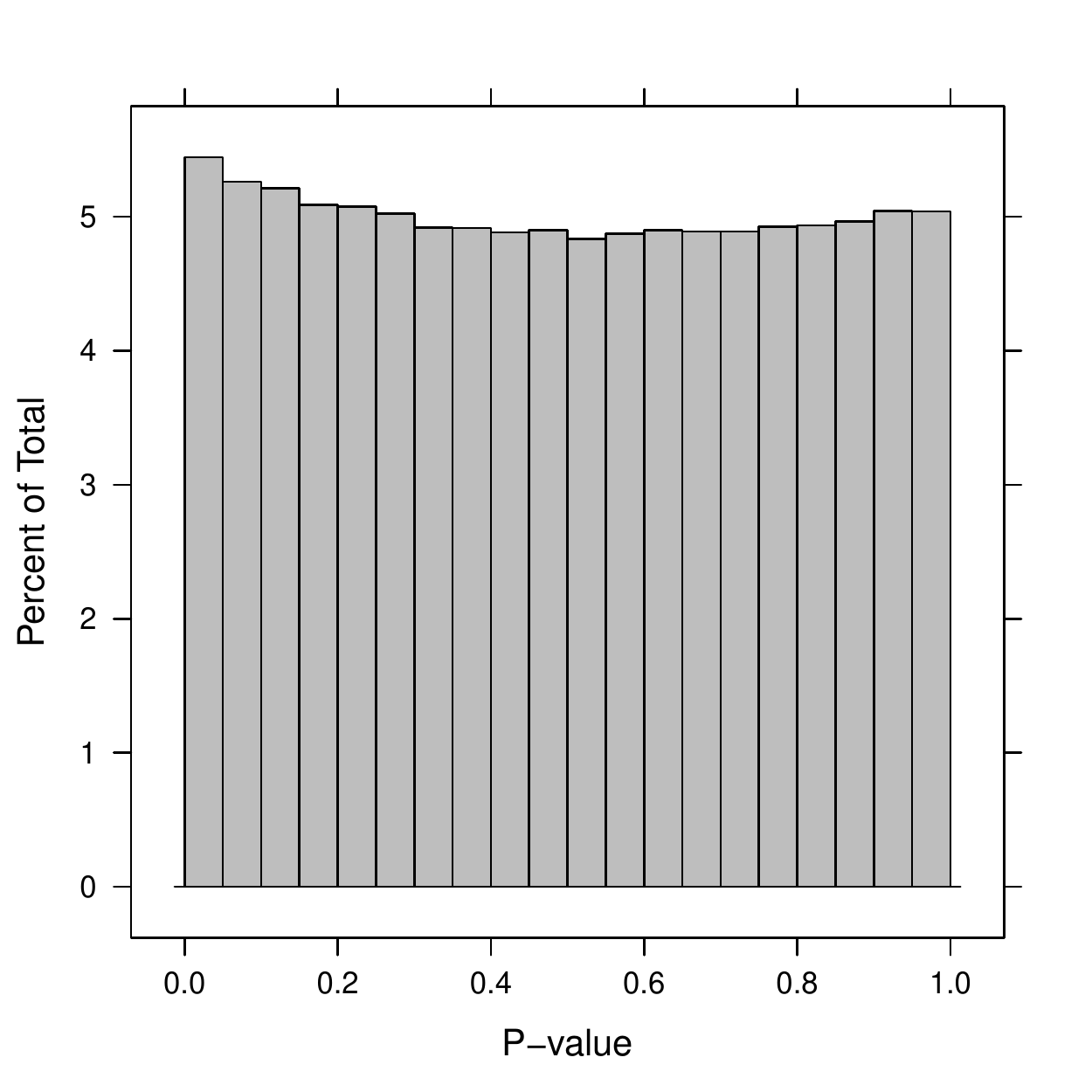}} 
\caption{\label{fig:null p-value} P-value histograms under the global null. We randomly place the 662 samples from Caucasian male living in far west into two equal-sized groups and produce their p-values for 5000 rounds. Top left is DM on the OTUs, top row right is DM on family level, bottom left is DM on class level, and bottom right is DTM .}
\end{figure}

Figure \ref{fig:null p-value} demonstrates the distribution of DM and DTM p-values under the global null. We fit three separate DM on 100 OTUs, family level and class level. The histogram of DTM p-values is produced by using all the p-values on the internal nodes. Surprisingly, the distribution of p-values for DM on the OTUs is far from being uniform on (0,1), which leads to conservative inference and loss of power. The discrepancy alleviates as we group more OTUs into family or class level, although its empirical distribution is still noticeably skewed. This phenomenon reflects the fact that DM is severely under-parametrized for microbiome data even in low dimensions, as it fits a single dispersion parameter that simultaneously controls all categories. In contrast, DTM solves this issue through fitting a family of dispersion parameters $\{\nu_A: A \in \mathcal{I}\}$ that leads to better calibrated p-values.

\begin{figure}[h]
\centering
\includegraphics[height=4.15cm]{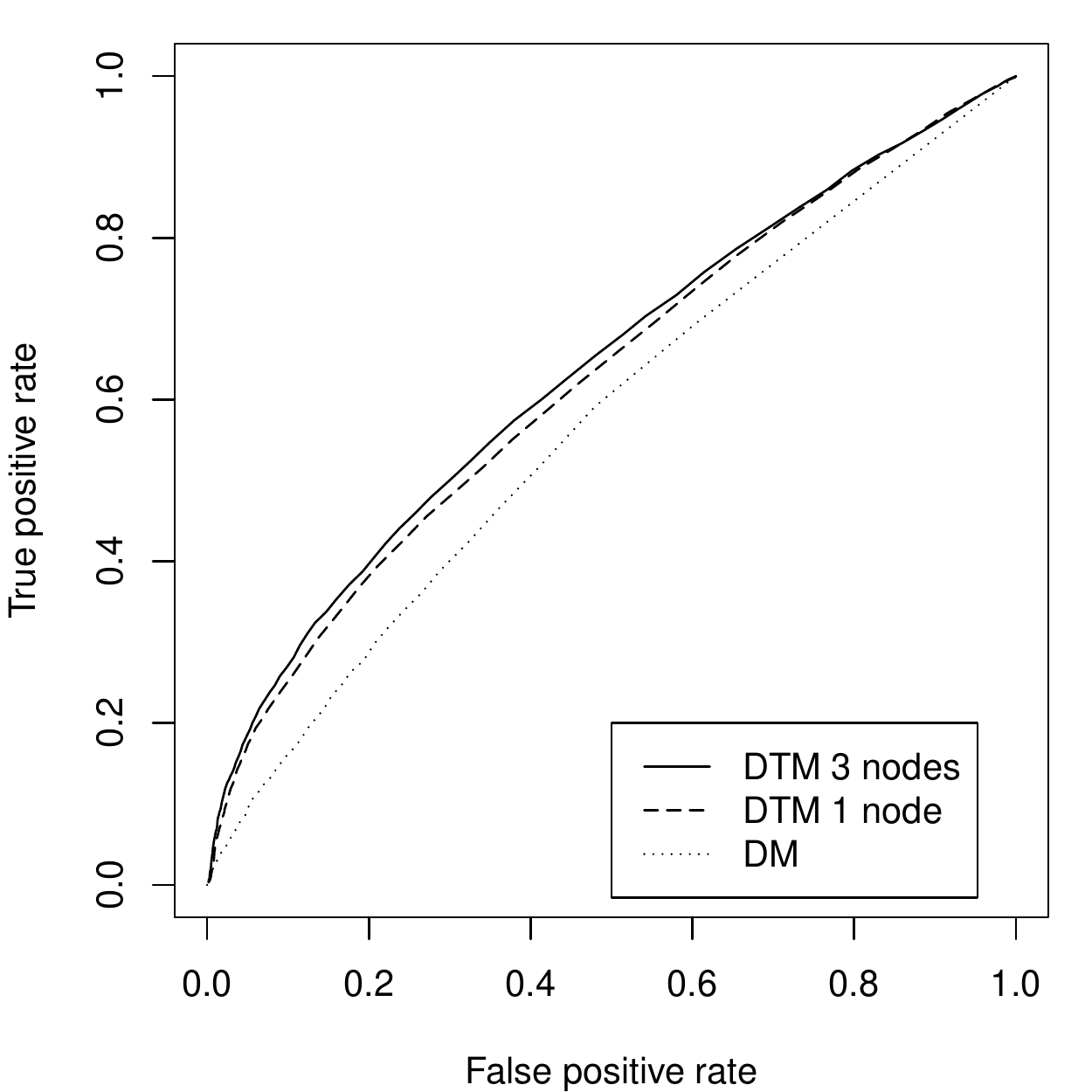}
\includegraphics[height=4.15cm]{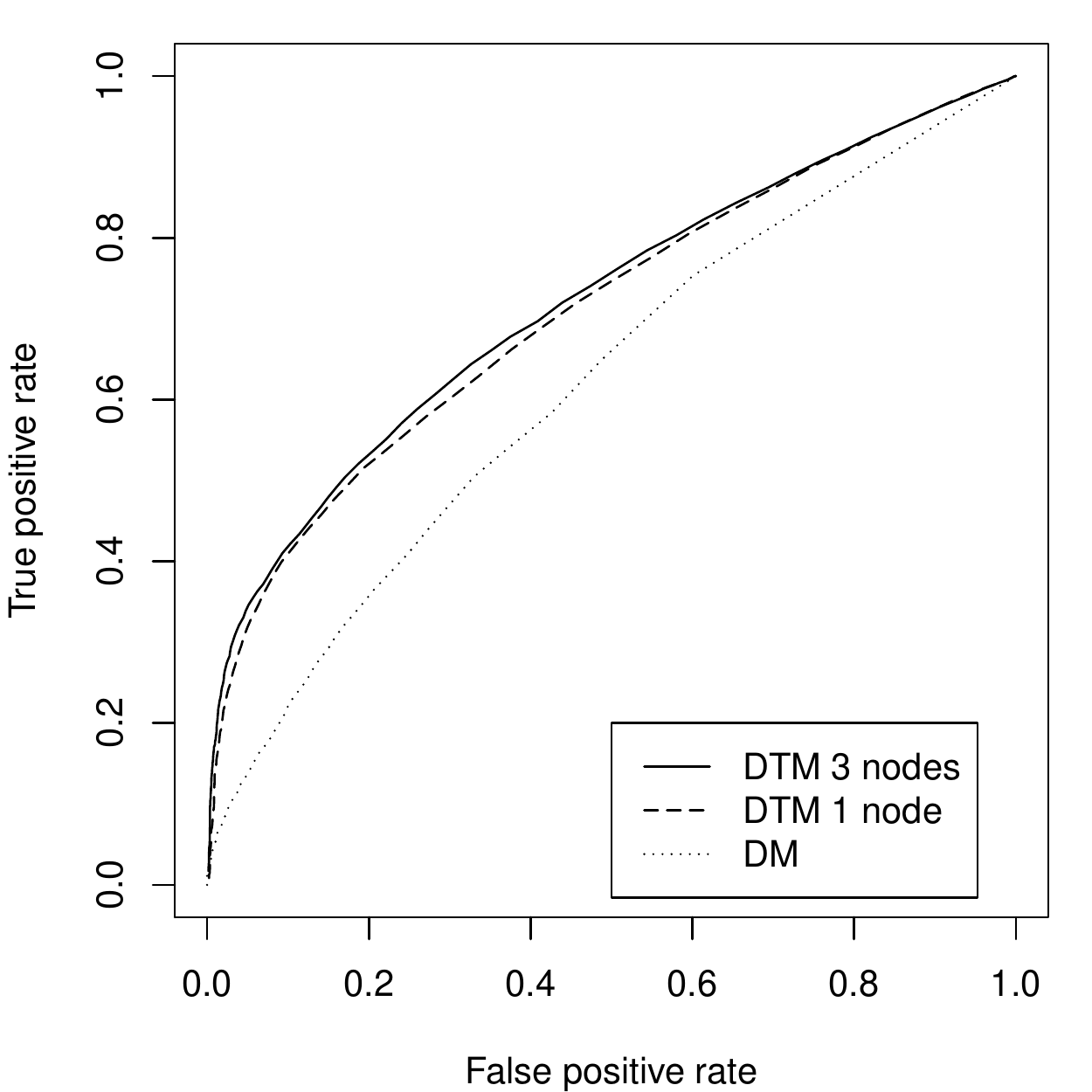}
\includegraphics[height=4.15cm]{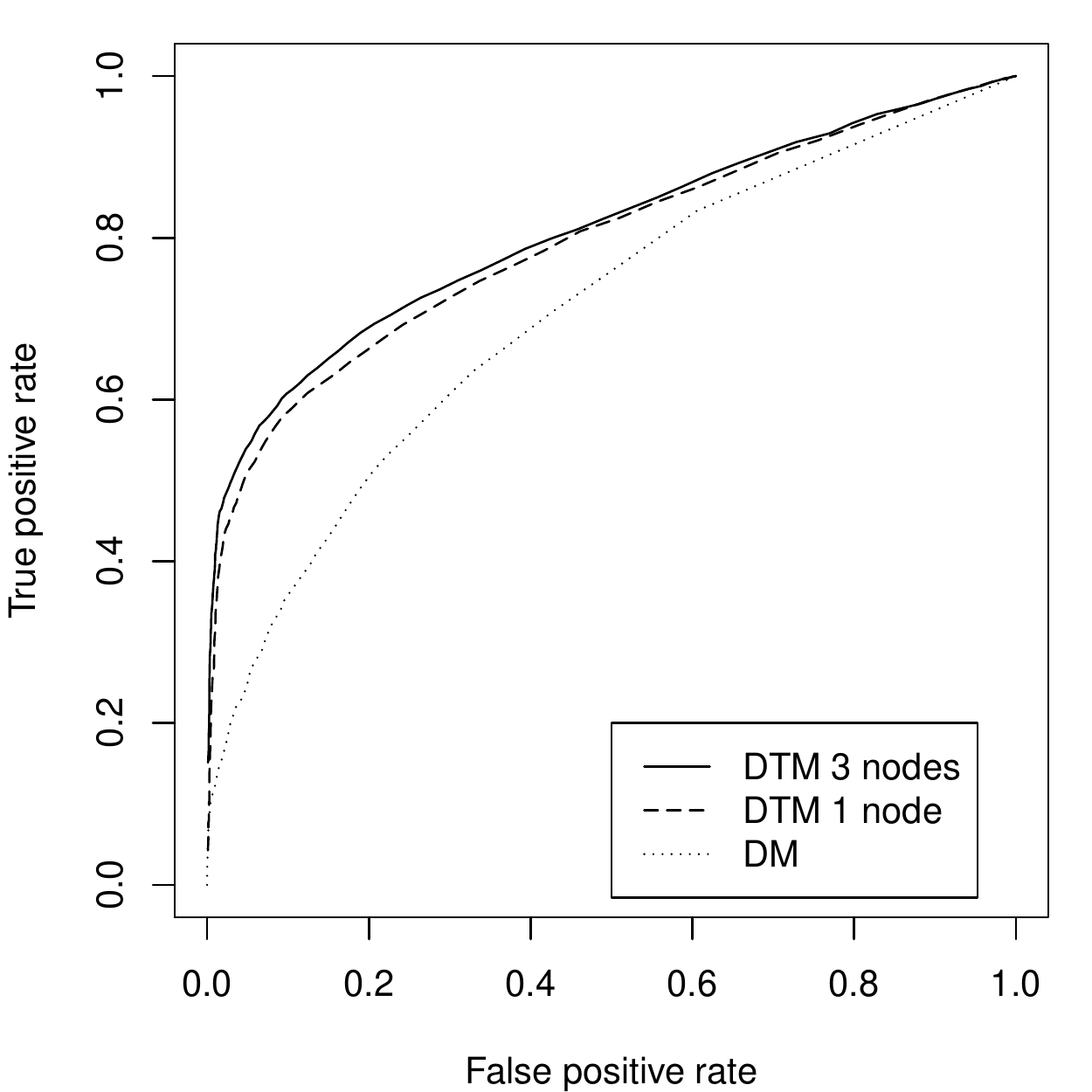}
\caption{\label{fig:ROC1} ROC curves from increasing the count of a random OTU. For left to right, the percentage increment is set as 100\%, 150\% and 250\%}
\end{figure}

\begin{figure}[h]
\centering
\includegraphics[height=4.15cm]{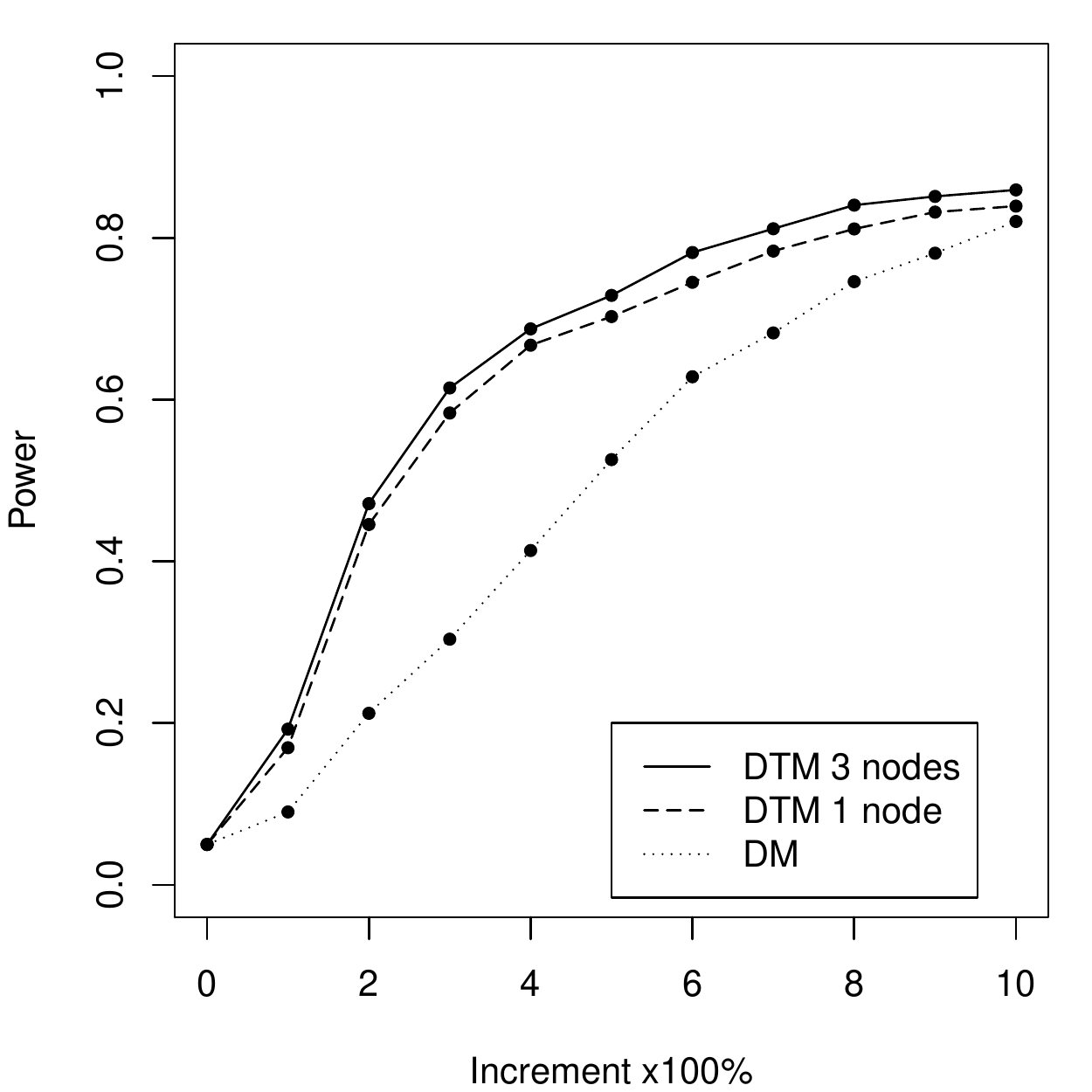}
\caption{\label{fig:power1} Power of DM and DTM with regard to different increment in a random OTU at false positive rate = 0.05.}
% * <marlee1982@gmail.com> 2016-10-15T17:10:07.786Z:
% 
% > false positive rate
% do you mean FWER? similarly for every place you mention ``false positive rate''
% 
% ^.
\end{figure}

Figure \ref{fig:ROC1} and \ref{fig:power1} show the ROC and power curves when we use the first simulation strategy to increase the count of a random OTU. We provide the result for (i) DM on the OTUs (ii) DTM using the maximum of the single node statistic, or $\max_{A \in \mathcal{I}} Z_A$ and (iii) DTM using the maximum of the triplet statistic, or $\max_{i \leq b} W_i$. The last strategy is the one employed in PhyloScan procedure. Both DTM methods give improved performance compared to DM due to highly localized signal.

\begin{figure}[h]
\centering
\subfloat{\includegraphics[height=4.15cm]{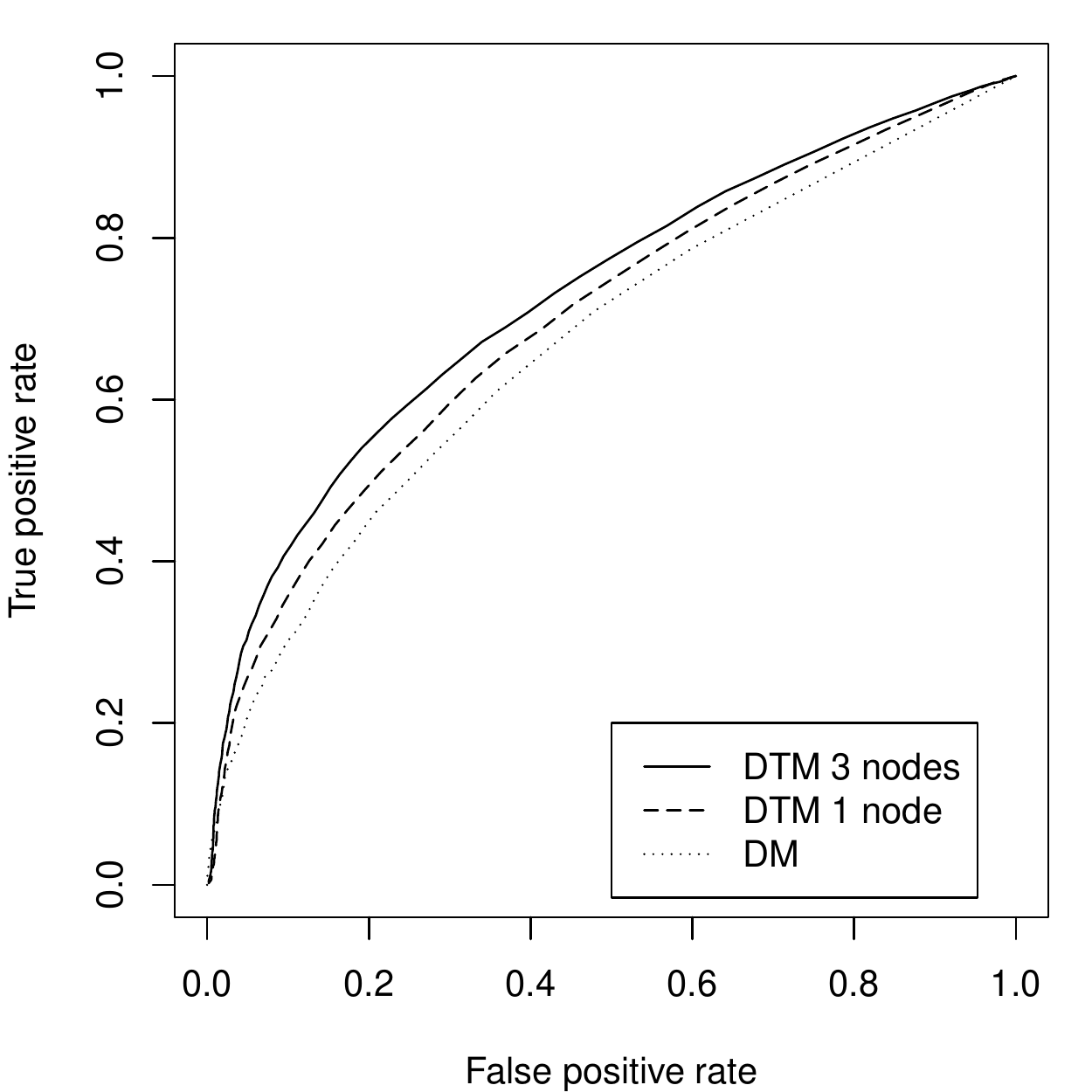}} 
\subfloat{\includegraphics[height=4.15cm]{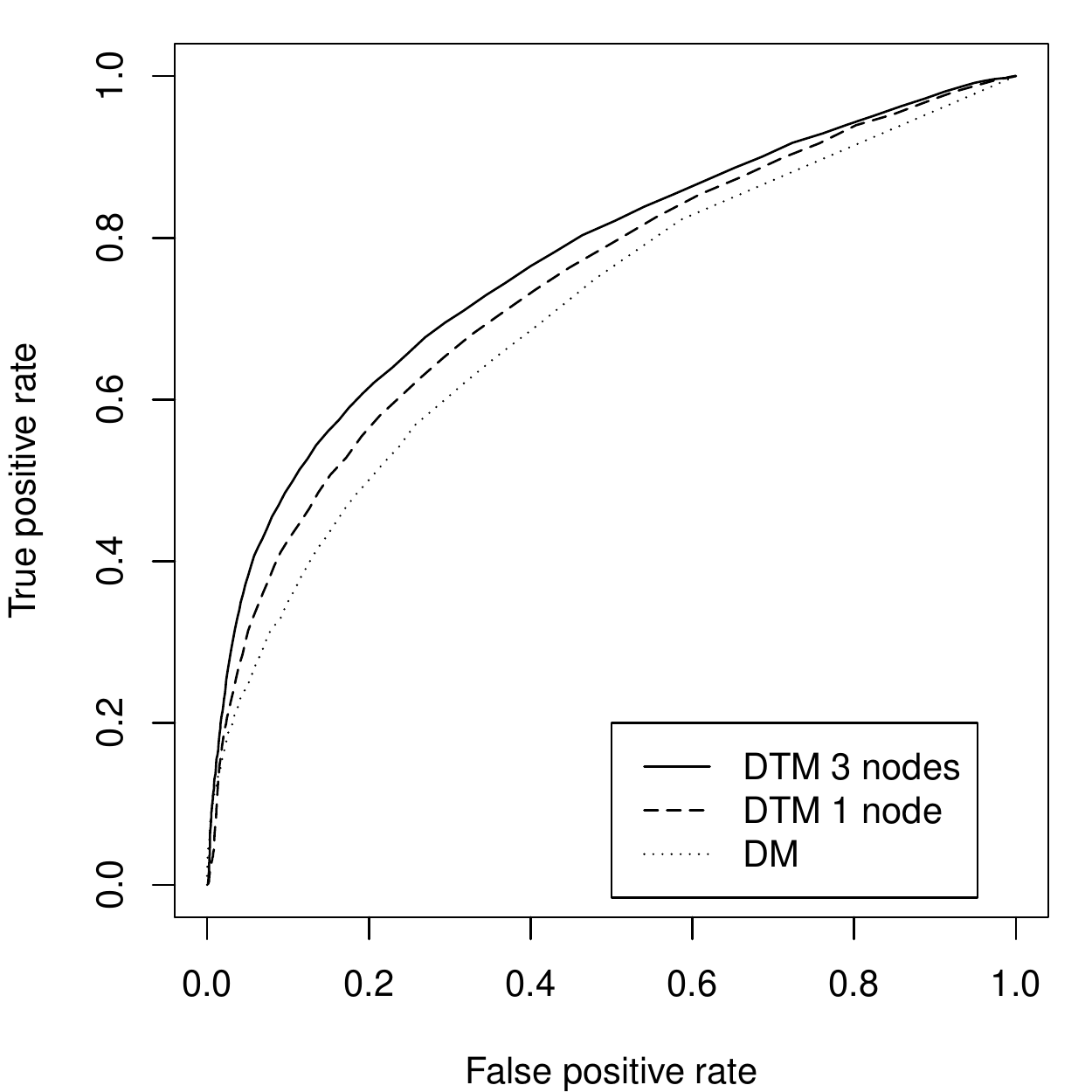}}
\subfloat{\includegraphics[height=4.15cm]{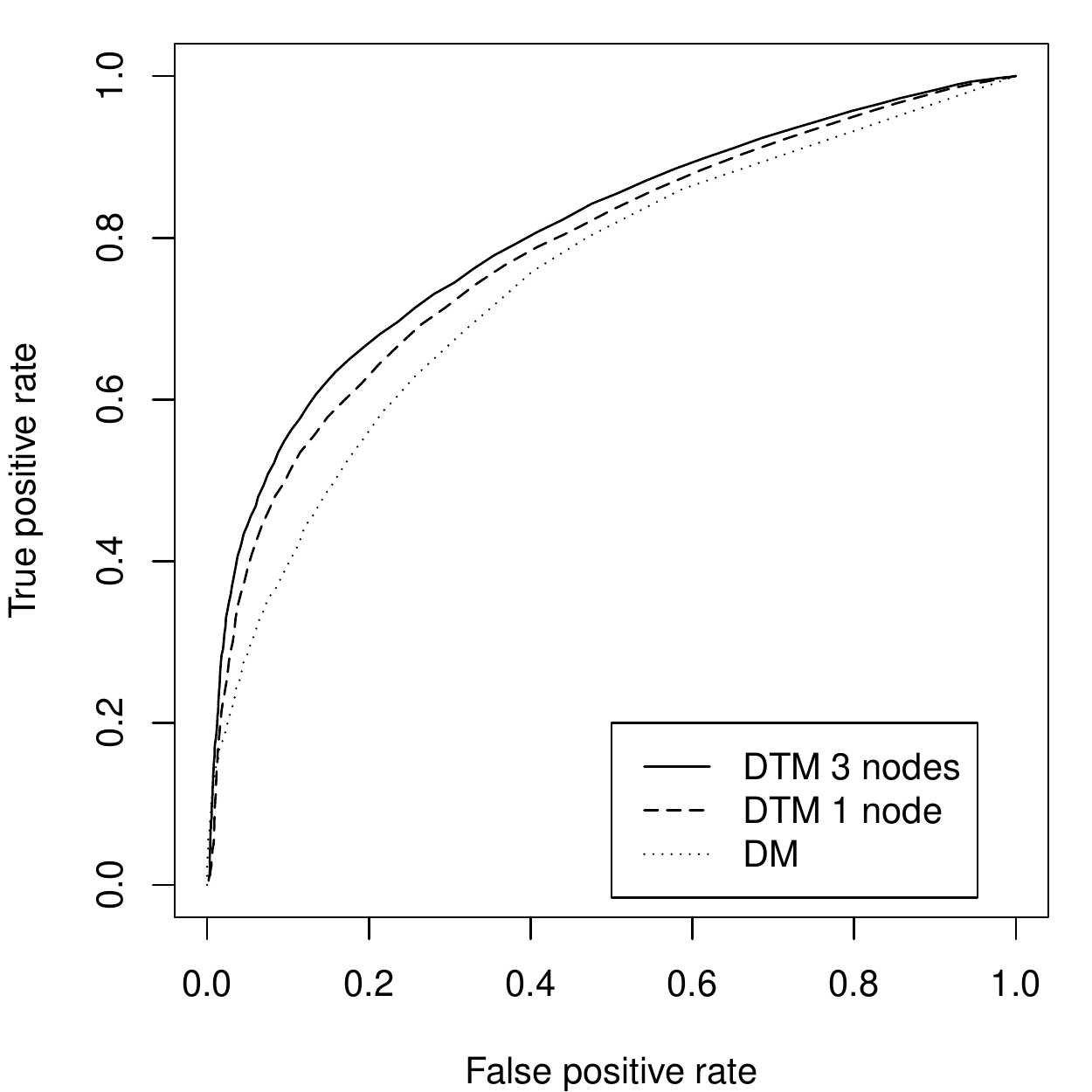}} \\ 
\subfloat{\includegraphics[height=4.15cm]{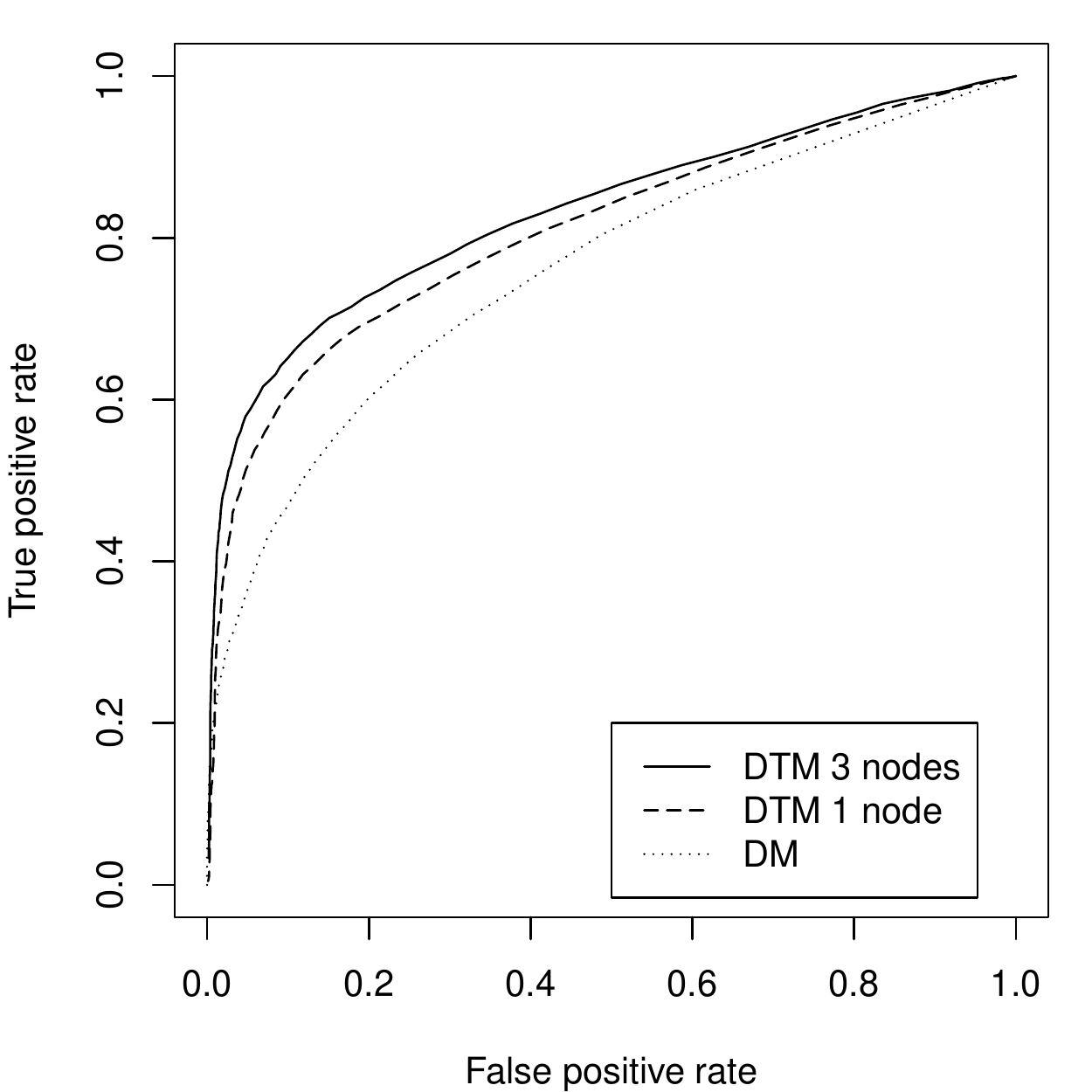}} 
\subfloat{\includegraphics[height=4.15cm]{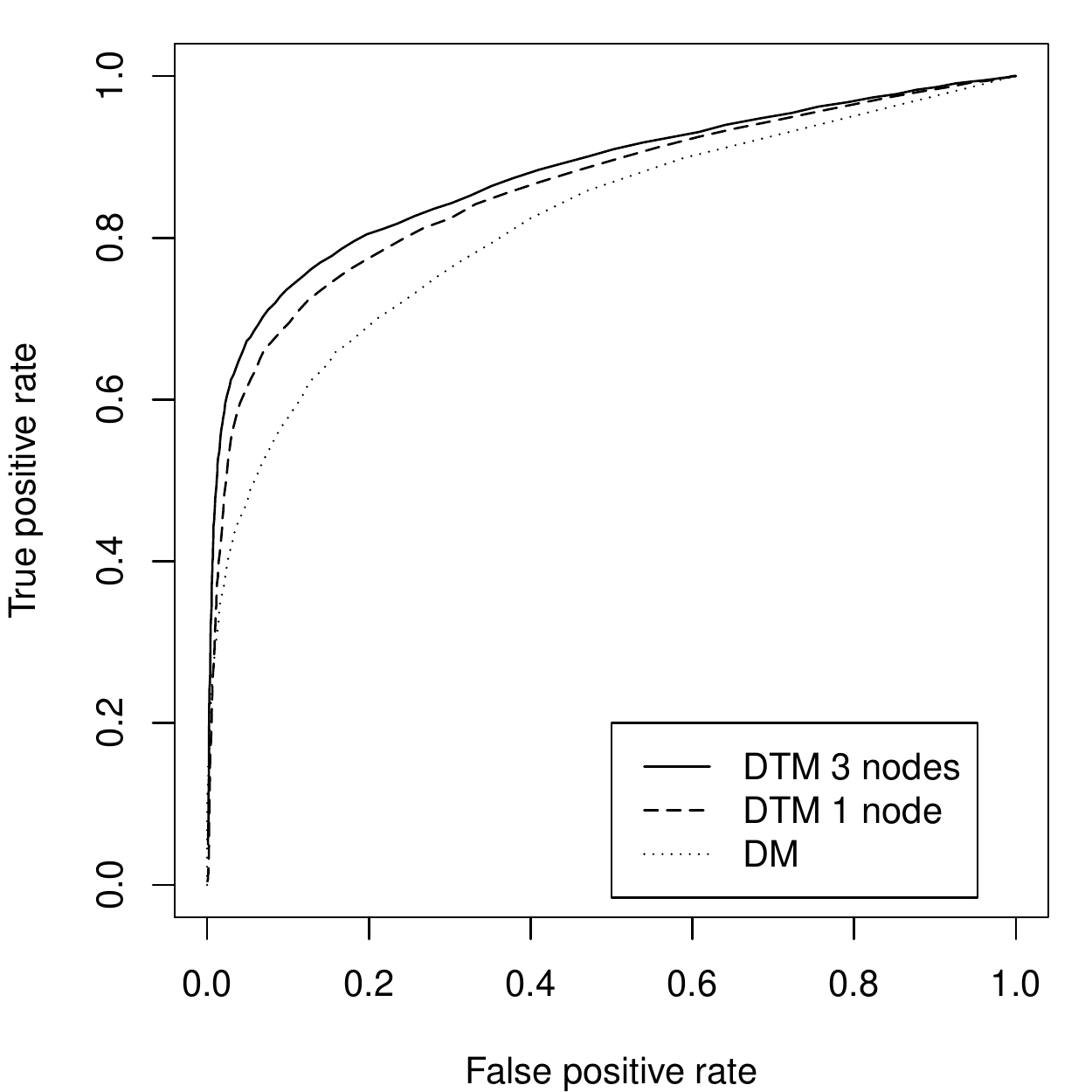}} 
\subfloat{\includegraphics[height=4.15cm]{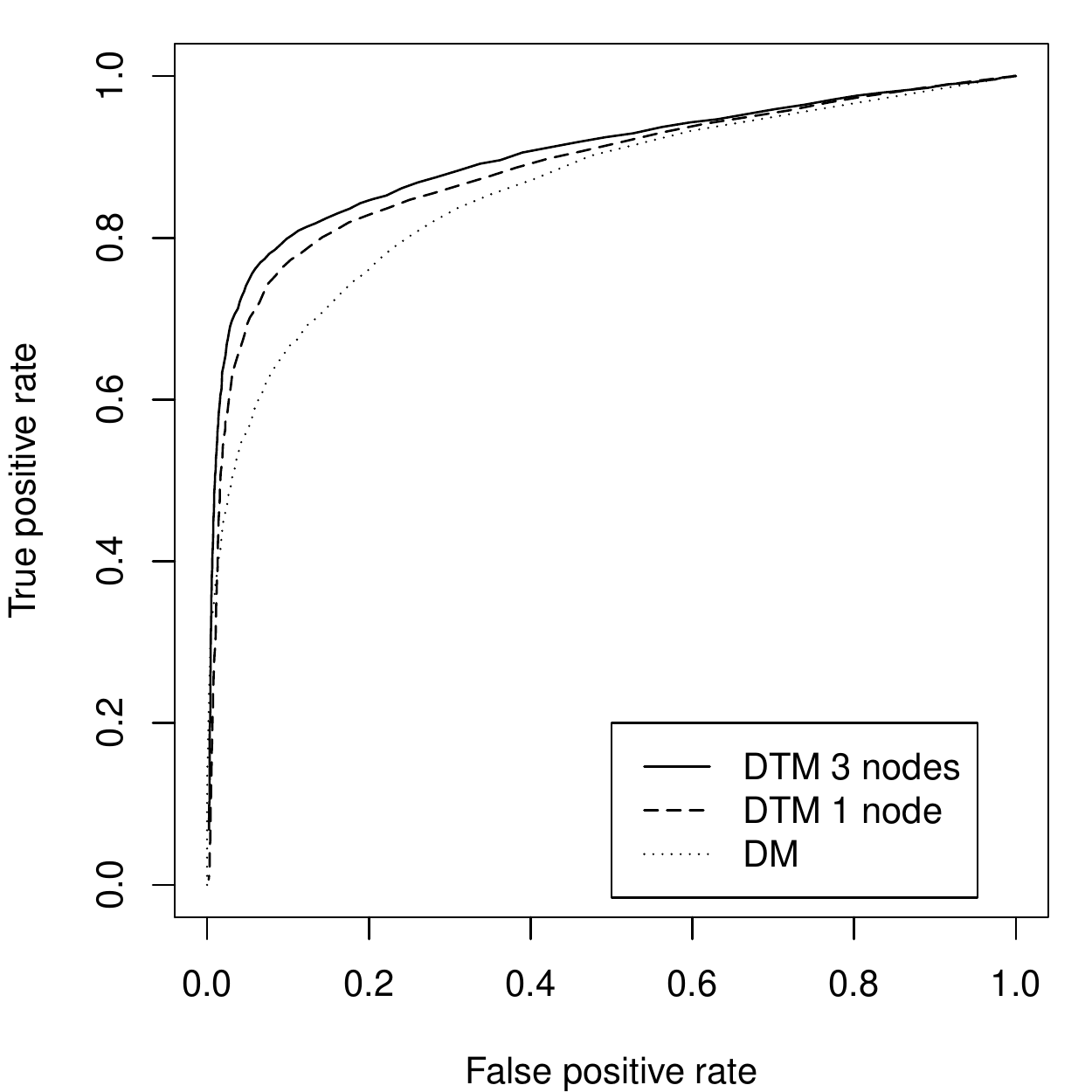}} \\
\caption{\label{fig:ROC2} ROC curves from increasing the count of all OTUs under a random internal node. The top row and the bottom row have the percentage increment set as 50\% and 75\% respectively. From left to right column, the minimum number of OTUs under the chosen internal node is 2, 3 and 5.}
\end{figure}

\begin{figure}[h]
\centering
\includegraphics[height=4.15cm]{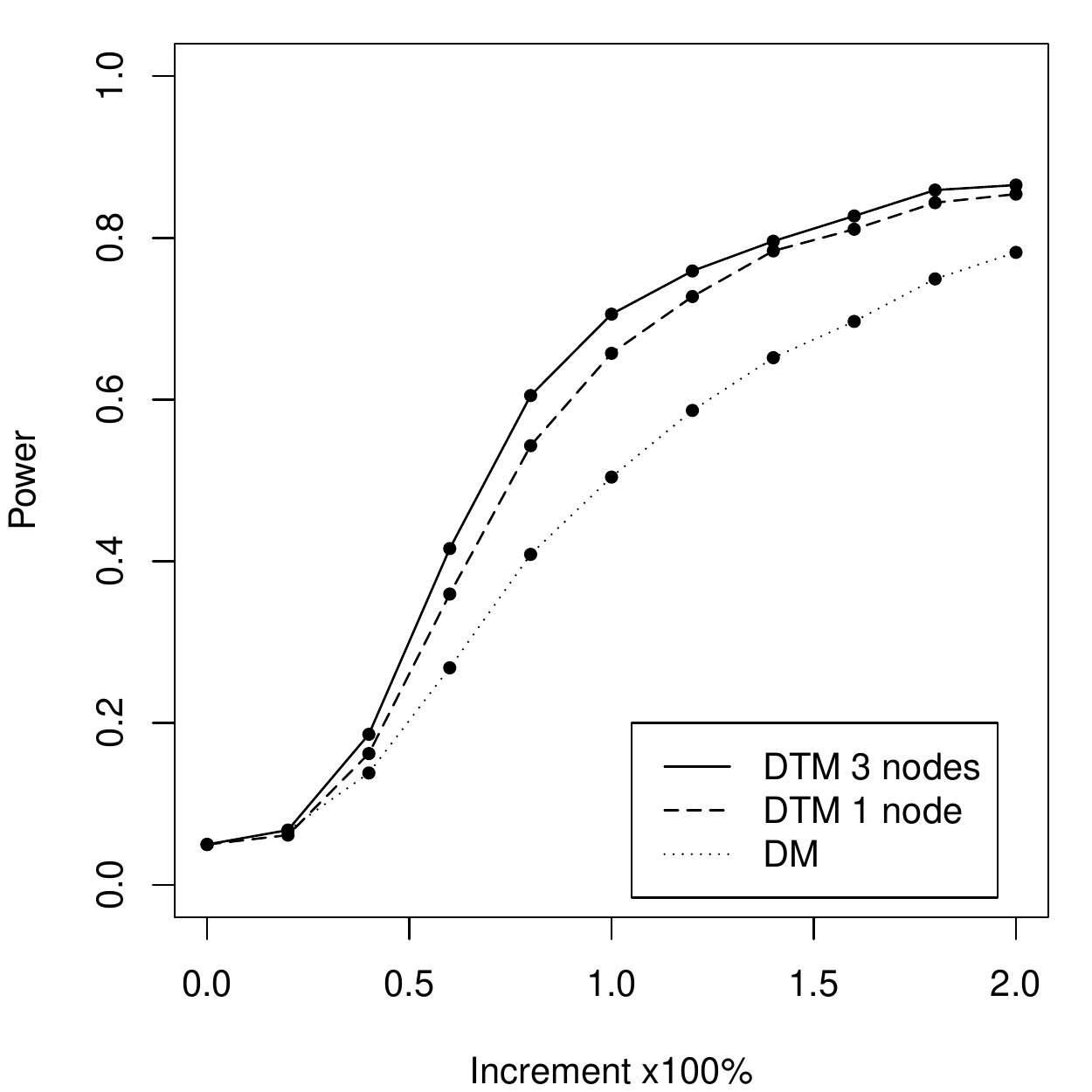}
\includegraphics[height=4.15cm]{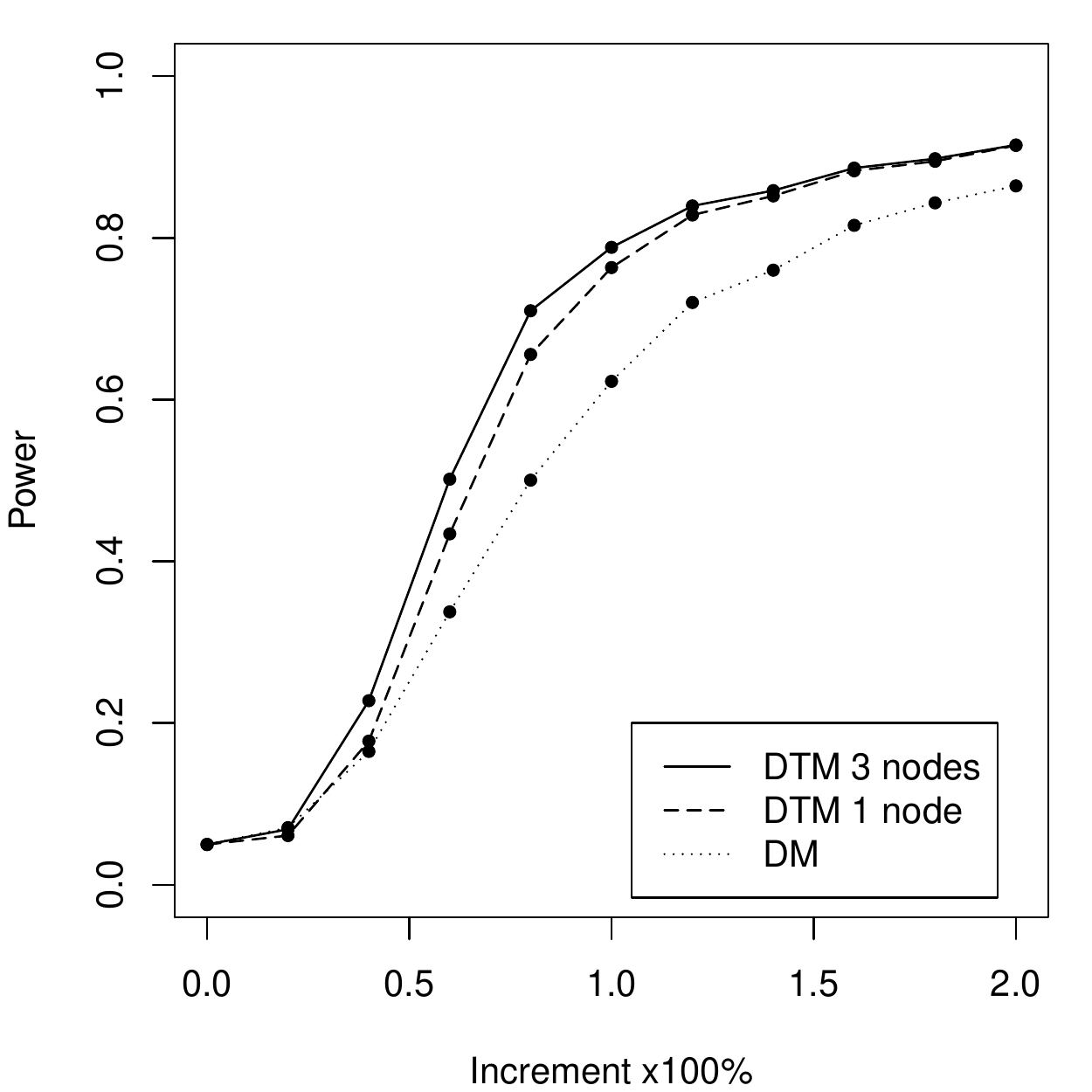}
\includegraphics[height=4.15cm]{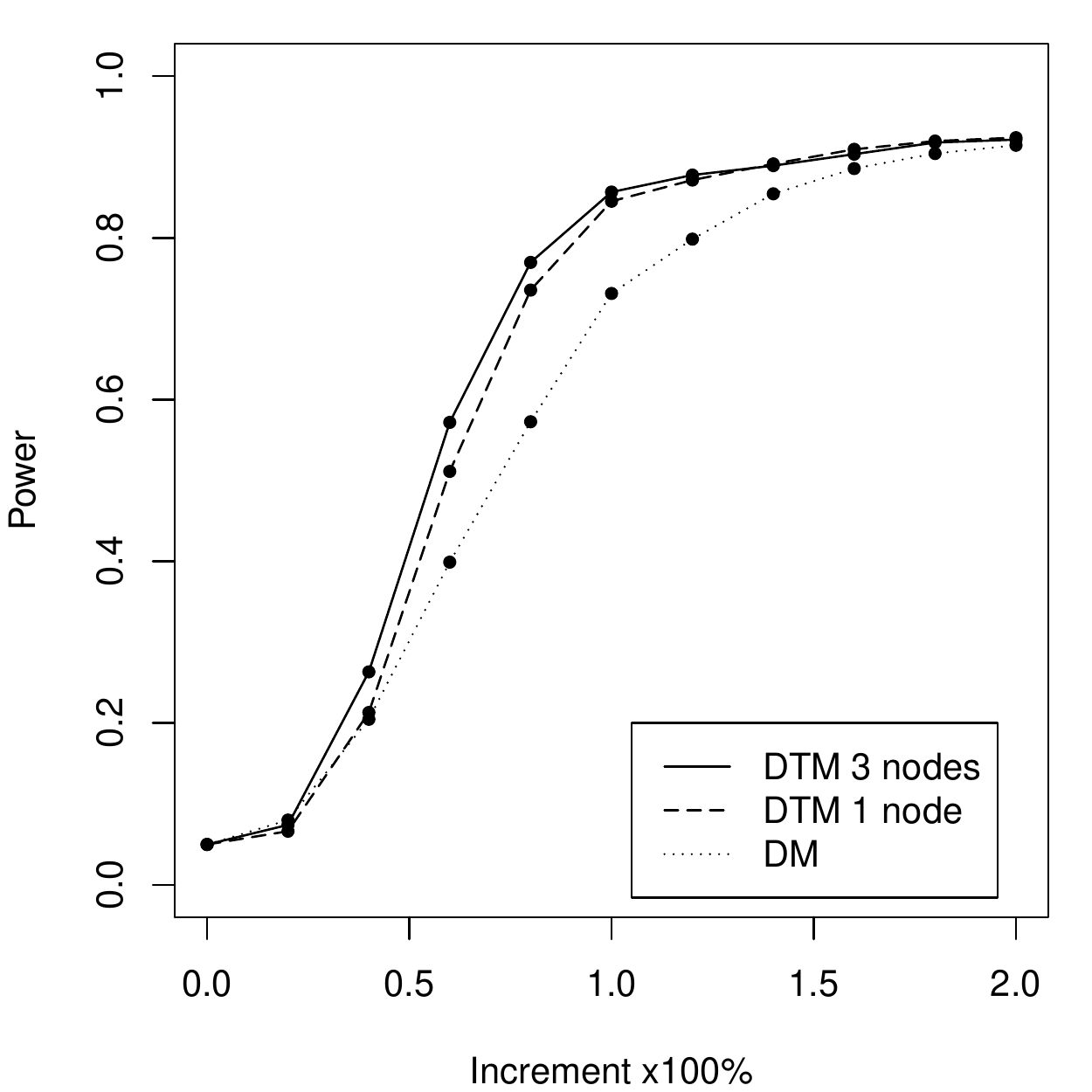}
\caption{\label{fig:power2} Power of DM and DTM with regard to different increment in all OTUs under a random internal node at false positive rate = 0.05. From left to right, the minimum number of OTUs under the chosen internal node is 2, 3 and 5.}
\end{figure}

Figure  \ref{fig:ROC2} and \ref{fig:power2} show the ROC and power curves when we use the second simulation strategy to increase the count of all OTUs under a random internal node. The minimum number of OTUs under the randomly selected internal node controls degree of localization in the signal. This simulation setup reflects the more biologically meaningful scenario in which a number of taxa exhibit differences in the between-group comparison. In all cases, DTM consistently provides higher power than DM. The DTM 3-node method also provides higher power than DTM 1-node at moderate increment levels. When the increment level is high, there will be a certain $Z_A$ whose value dominates all other node statistics, so the extra gain from pooling signal strength within triplets diminishes.
 
\section{Discussion} \label{sec:Discussion}

DTM models the microbiome data through a cascade of local DMs with varying degrees of resolutions on the phylogenetic tree. We take advantage of the correlated signals on the tree through a scan statistic approach and provide upper and lower bound on its tail probability for testing cross-group differences. Both empirical results on American Gut data and simulations demonstrated the efficiency and accuracy of our method. We also developed the PhyloScan R package, which can be found at \href{https://github.com/yunfantang/PhyloScan}{https://github.com/yunfantang/PhyloScan}.

DTM is a generalization of DM with $|\mathcal{I}| - 1$ more dispersion parameters. An interesting question is whether one could stepwise tune the model (hence the number of parameters) from DM to DTM. To start, the DTM representation in (\ref{PhyloDM0}) shrinks to the degenerate DM if $\exists \nu>0$ s.t. $ \nu_A = \nu\sum_{w \in A} \pi_w$ for all $A \in \mathcal{I}$. This condition is equivalent to $\nu_A = \nu_{R(A)}\pi_{R(A),i}$ for $\forall A \in \mathcal{I}\backslash\{\Omega\}$ with $A = \mathcal{C}(R(A))_i$. Stepwise tuning can be achieved through requiring only $\nu_A = \nu_{R(A)}\pi_{R(A),i}$ to hold over $A \in \tilde{\mathcal{I}}$ where $\tilde{\mathcal{I}} \subset \mathcal{I}\backslash\{\Omega\}$ controls the effective degrees of freedom. Apparently $\tilde{\mathcal{I}} = \emptyset$ leads to DTM and $\tilde{\mathcal{I}} = \mathcal{I}\backslash\{\Omega\}$ leads to DM, so any choice of $\tilde{\mathcal{I}}$ in the middle yields a model between the two extremes. Standard model selection techniques such as information criterion or cross validation can then be applied.
Although the existence of such spectrum grants substantial  flexibility, we note that it can be computationally infeasible to examine the model fit of all $2^{|\mathcal{I}|-1}$ possible configurations. A potential workaround is to enlarge $\mathcal{T}$ stepwise by a greedy algorithm or use dynamic programming, but it is not clear under which conditions we are guaranteed to recover the global optimum. 

Since our PhyloScan procedure only requires p-value as input, it can be easily applied to any extensions or other distributions. For example, the DTM framework can be adapted to incorporate continuous variable of interest and adjust for the effects of confounders. When the tree is fully binary, we let $\lambda_A = \pi_{A,1}$ to fully represent $\boldsymbol{\pi}_A = (\pi_{A,1}, 1-\pi_{A,1})$ in (\ref{PhyloDM0}). Then we can build separate logistic regression models for each $A$:
\begin{equation}
\log\frac{\lambda_A}{1-\lambda_A} = \beta_{A,0} + \beta_{A,1}u + \sum_{i=1}^{s}\beta_{A,s+1}c_s
\end{equation}
where $\beta_{A,i}$ is the $i$th regression coefficient, $u$ denotes the continuous variable of interest and $c_1$, $c_2$, ..., $c_s$ are the confounders. After obtaining maximum likelihood estimates of the coefficients as well as $\nu_A$, we test the significance of $u$'s coefficient to produce p-values and use them as input to PhyloScan  in order to borrow strength from neighboring nodes. Another possible extension is related to the issue of zero-adjustment. In Figure \ref{fig:null p-value}, DM p-values exhibit apparent right skew under the global null hypothesis. A follow-up inspection shows that MoM estimation tends to produce higher expected zero count than observed. Both right-skewness and zero-deflation of the global DM are likely caused by underestimation of $\nu$, which makes the Dirichlet prior more dispersed. In DTM, we still observe mild level of zero-deflation, although the extent is much less severe than global DM. It is also possible to have zero-inflation when one switches to a different sequencing technology or OTU construction algorithm. Incorporating zero-adjustment into existing model can lead to significantly better fit while easily handled by PhyloScan.

\section*{Acknowledgements} L. Ma is partly supported by NSF grant DMS-1612889 and a Google Faculty Research Award. D. L. Nicolae is supported in part by NIH grants R01-MH101820 and R01-HL129735. Part of the research was completed when L. Ma was a Visiting Scholar in the Department of Statistics at University of Chicago in 2016.

\appendix
\section*{Appendix}
\subsection*{A1. Proof of Theorem 1} The elements in $\mathcal{I}$ can be ordered as $A_{1}, A_{2}, ..., A_{|\mathcal{I}|}$ such that each parent node always appears in front of its children. Let $p_{A_{l}}$ be the p-value for testing $H_{0,A_{l}}$. Without loss of generality, let us assume $A_{|\mathcal{I}|} = \{K-1, K\}$. For any subject with OTU counts $\boldsymbol{x}$, the probability function (\ref{PhyloDM2}) can be written as
\begin{align*}
f_{T}(\boldsymbol{x}) &= \prod_{l=1}^{\mathcal{I}} f\big(\boldsymbol{x}(A_{l})|N(A_{l})\big) \\
&= \prod_{l=1}^{\mathcal{I}-1} f\big(\boldsymbol{x}(A_{l})|N(A_{l})\big) \cdot f\big(\boldsymbol{x}(A_{|\mathcal{I}|})|N(A_{|\mathcal{I}|})\big) \\
&= f_T\big(x_1, x_2, ..., x_{K-2}, N(A_{|\mathcal{I}|})\big)  f\big(\boldsymbol{x}(A_{|\mathcal{I}|})|N(A_{|\mathcal{I}|})\big),
\end{align*}
which yields
\begin{equation*}
f_{T}\big(\boldsymbol{x} | N(A_{|\mathcal{I}|}) \big) = f_T\big(x_1, x_2, ..., x_{K-2}, N(A_{|\mathcal{I}|}) | N(A_{|\mathcal{I}|})\big)  f\big(\boldsymbol{x}(A_{|\mathcal{I}|})|N(A_{|\mathcal{I}|})\big).
\end{equation*}
Since the above conditional independence relationship holds for all subjects, it follows that $p_{A_{|\mathcal{I}|}}$ is independent of all other $p_{A_{l}}$'s conditional on $N(A_{|\mathcal{I}|})$. Therefore under $H_0$,
\begin{align*}
P(\bigcup_{l=1}^{|\mathcal{I}|} \{p_{A_{l}} \leq \alpha_l\} ) &= E\Big(  P \big( \bigcup_{l=1}^{|\mathcal{I}|} \{p_{A_{l}} \leq \alpha_l\} | N(A_{|\mathcal{I}|}) \big) \Big)  \\
&= E\Big(  P\big(\bigcup_{l=1}^{|\mathcal{I}|-1} \{p_{A_{l}} \leq \alpha_l\} | N(A_{|\mathcal{I}|})\big) P\big(p_{A_{|\mathcal{I}|}}\leq\alpha_{|\mathcal{I}|} | N(A_{|\mathcal{I}|})\big)  \Big) \\
&= \alpha_{|\mathcal{I}|} E\Big(  P\big(\bigcup_{l=1}^{|\mathcal{I}|-1} \{p_{A_{l}} \leq \alpha_l\} | N(A_{|\mathcal{I}|})\big) \Big) \\
&= \alpha_{|\mathcal{I}|} P(\bigcup_{l=1}^{|\mathcal{I}|-1} \{p_{A_{l}} \leq \alpha_l\} ).
\end{align*} 
where the second last equation requires the asymptotic distribution of (\ref{DMtesting}) holds so that $P\big(p_{A_{|\mathcal{I}|}}\leq\alpha_{|\mathcal{I}|} | N(A_{|\mathcal{I}|})\big) = \alpha_{|\mathcal{I}|}$

Repeating the above procedures iteratively for $A_{|\mathcal{I}|-1}$, $A_{|\mathcal{I}|-2}$, ... $A_{1}$ gives
$$P(\bigcup_{l=1}^{|\mathcal{I}|} \{p_{A_{l}} \leq \alpha_l\} ) = \prod_{l=1}^{|\mathcal{I}|} \alpha_l.$$

\subsection*{A2. Proof of Theorem 2} By (\ref{UP4}),
\begin{align*}
\epsilon_U &\leq \sum_{i=1}^b\sum_{j<i, j\notin \mathcal{N}_i} P(B_i \cap B_j  \cap M^c ) \\
&= \sum_{i\leq b, \mathcal{B}_i \notin \mathcal{M}}\sum_{j<i, j\notin \mathcal{N}_i, \mathcal{B}_j \notin \mathcal{M}} P(B_i \cap B_j  \cap M^c ).
\end{align*}
The elements inside the summation sign above fall in one of the following two categories
\begin{enumerate}
\item[(i)] $|\mathcal{B}_i \cap \mathcal{B}_j| = 0$ which means $P(B_i \cap B_j \cap M^c) < P(B_i \cap B_j) = \big(1-F_3(w)\big)^2$ 
\item[(ii)] $|\mathcal{B}_i \cap \mathcal{B}_j| = 1$ so that $P(B_i \cap B_j \cap M^c) \leq P(\sum_{i=1}^3 Y_i>w,  \sum_{i=3}^5 Y_i>w, Y_i < w \text{ for all } i)$, where $Y_i$'s are i.i.d. chi-square distributed with 1 degree of freedom.
\end{enumerate}
Let $f_1(y) = \frac{y^{-\frac{1}{2}}e^{-\frac{y}{2}}}{\sqrt{2\pi}}$ be the density function of $\chi^2_1$. Conditioning on $Y_3$ gives the following upper bound on category (ii):
\begin{align*}
P(\sum_{i=1}^3 Y_i>w,  \sum_{i=3}^5 Y_i>w, Y_i < w) &< \int_0^w f_1(y)P(\sum_{i=1}^3 Y_i>w|Y_3=y) \cdot \\
&\hspace{5mm} P(\sum_{i=3}^5 Y_i>w|Y_3=y) dy \\ 
&= \int_0^w \frac{y^{-\frac{1}{2}}e^{-\frac{y}{2}}} {\sqrt{2\pi}} \big(1-F_2(w-y)\big)^2 dy \\ 
&= \frac{1}{\sqrt{2\pi}} \int_0^w y^{-\frac{1}{2}}e^{-\frac{y}{2}} e^{-(w-y)} dy \\
&= \frac{e^{-w}}{\sqrt{2\pi}} \int_0^w y^{-\frac{1}{2}}e^{\frac{y}{2}}dy \\
&< 0.9w^{-\frac{1}{2}}e^{-\frac{w}{2}} \text{ for } w \geq 12,
\end{align*}
where the last line is deduced by noticing that the function $h(w) = \int_0^w y^{-\frac{1}{2}}e^{\frac{y}{2}}dy - 2.25w^{-\frac{1}{2}}e^{\frac{w}{2}}$ satisfies 1) $h(12) < 0$ and 2) $h'(w) = -0.125w^{-\frac{1}{2}}e^{\frac{w}{2}} + 1.125w^{-\frac{3}{2}}e^{\frac{w}{2}} = e^{\frac{w}{2}}\cdot$ $w^{-\frac{1}{2}}(1.125w^{-1}-0.125) < 0 $ for $w \geq 12$. Together they establish $\int_0^w y^{-\frac{1}{2}}e^{\frac{y}{2}}dy < 2.25w^{-\frac{1}{2}}e^{\frac{w}{2}}$ for $w \geq 12$. 

Since there are $\xi_1$ terms in (i) and $\xi_2$ terms in (ii), we have 
\begin{equation}
\epsilon_U < \xi_1\big(1-F_3(w) \big)^2 + 0.9\xi_2w^{-\frac{1}{2}}e^{-\frac{w}{2}}. \label{Thm2_0}
\end{equation}

Our next step is to put a lower bound on $P_U$. According to (\ref{UP3}),
\begin{align}
P_U &= P(M) + \sum_{i=1}^b P(B_i \cap B_{\mathcal{N}_i}^c \cap M^c ) \label{Thm2_1}.
\end{align}

The lower bound of first term in the right side of (\ref{Thm2_1}) is obtained by considering only the triplets in $\mathcal{M}$:
\begin{align}
P(M) &\geq 1 - F_3(w)^{\xi_3} \nonumber \\
&= 1 - \Big(1 - \big(1- F_3(w)\big)\Big)^{\xi_3} \nonumber \\
&\geq \xi_3 \big(1-F_3(w)\big) - \frac{\xi_3(\xi_3-1)}{2}(1-F_3(w)\big)^2 \text{, by Taylor expansion} \nonumber \\
&\geq 0.95\xi_3\big(1-F_3(w)\big), \text{ as long as } \xi_3\big(1-F_3(w)\big) < 0.1. \label{Thm2_1.5} 
\end{align}

Next, we get the lower bound of the second term in the right side of (\ref{Thm2_1}). For any fixed $i$, let $Y_1, Y_2$ and $Y_3$ denote the i.i.d. $\chi^2_1$ variables included in the event $B_i$ so that $B_i = \{\sum_{j=1}^3 Y_j >w\}$. Then we have
\begin{align*}
P(B_i \cap B_{\mathcal{N}_i}^c \cap M^c ) &= \int_{\sum y_i>w}f_1(y_1)f_1(y_2)f_1(y_3)P(B_{\mathcal{N}_i}^c \cap M^c  | y_1,y_2,y_3) d\boldsymbol{y} \\
&= \int_{\sum y_i>w}f_1(\boldsymbol y)h(\boldsymbol y ,w) d\boldsymbol{y}
\end{align*}
where we define $f_1(\boldsymbol y) = f_1(y_1)f_1(y_2)f_1(y_3)$ and $h(\boldsymbol y ,w) = P(B_{\mathcal{N}_i}^c \cap M^c  | y_1,y_2,y_3)$. In addition, let $V(w) = \{\boldsymbol y: \sum_{i=1}^3 y_i >w\}$ denote the region of integration. By Reynolds transport theorem (or multidimensional Leibniz's rule),
\begin{align}
\frac{d}{dw} P(B_i \cap B_{\mathcal{N}_i}^c \cap M^c ) &= \int_{V(w)}f_1(\boldsymbol y)\frac{\partial}{\partial w} h(\boldsymbol y ,w)d\boldsymbol y  \nonumber \\ 
&\hspace{5mm} + \int_{\partial V(w)} (\boldsymbol v_b \cdot \boldsymbol n) f_1(\boldsymbol y)h(\boldsymbol y ,w) dA  \label{Thm2_2}
\end{align}
where $\partial V(w)$ is the boundary of $V(w)$, $\boldsymbol v_b$ is the Eulerian velocity of the boundary, $\boldsymbol n$ is the outward-pointing unit-normal, and $dA$ is the surface element.

On the other hand,
\begin{align}
\frac{d}{dw} P(B_i) &= \frac{d}{dw} \int_{V(w)} f_1(\boldsymbol y) d\boldsymbol y \nonumber \\ 
&= \int_{\partial V(w)} (\boldsymbol v_b \cdot \boldsymbol n) f_1(\boldsymbol y) dA \label{Thm2_3}
\end{align}

Since both $B_{\mathcal{N}_i}^c$ and $M^c$ strictly enlarges as $w$ increases, $h(\boldsymbol y ,w)$ is an increasing function of $w$. This leads to $\frac{\partial}{\partial w}h(\boldsymbol y ,w) > 0$. Moreover, $h(\boldsymbol y ,w) < 1$ from definition. Lastly, the fact that $V(w_1) \subset V(w_2)$ for any $w_1 > w_2$ gives $\boldsymbol v_b \cdot \boldsymbol n \leq 0$. These altogether establish $\frac{d}{dw} P(B_i \cap B_{\mathcal{N}_i}^c \cap M^c ) >\frac{d}{dw} P(B_i)$ as we compare the expression in $(\ref{Thm2_2})$ and $(\ref{Thm2_3})$. With the apparent relation $P(B_i \cap B_{\mathcal{N}_i}^c \cap M^c ) < P(B_i)$, we conclude that
\begin{equation}
\frac{d}{dw} \cdot \frac{P(B_i \cap B_{\mathcal{N}_i}^c \cap M^c )}{P(B_i)} > 0 \text{ for all }  i \label{Thm2_4}
 \end{equation}

Therefore if we only focus on calculating $\epsilon_U/P_U$ for $w \geq w_T$ where $w_T$ is a pre-fixed value, then we can first evaluate $\sum_{i=1}^b P(B_i \cap B_{\mathcal{N}_i}^c \cap M^c ) / \big(1-F_3(w)\big) = \xi_T$ at $w = w_T$. By (\ref{Thm2_4}),
\begin{equation}
\sum_{i=1}^b P(B_i \cap B_{\mathcal{N}_i}^c \cap M^c ) \geq \xi_T \big(1-F_3(w)\big) \text{ for } w \geq w_T \label{Thm2_5}
\end{equation}
Plugging in (\ref{Thm2_1.5}) and (\ref{Thm2_5}) into (\ref{Thm2_1}) gives
\begin{equation}
P_U \geq (0.95\xi_3 + \xi_T) \big(1-F_3(w)\big) \label{Thm2_6}
\end{equation}

The upper bound on $\epsilon_U$ in (\ref{Thm2_0}) and the lower bound on $P_U$ in (\ref{Thm2_6}) yield
\begin{align}
\frac{\epsilon_U}{P_U} &< \frac{\xi_1\big(1-F_3(w) \big)^2 + 0.9\xi_2w^{-\frac{1}{2}}e^{-\frac{w}{2}}}{(0.95\xi_3 + \xi_T) \big(1-F_3(w)\big)} \nonumber \\
&= \frac{\xi_1\big(1-F_3(w) \big)}{0.95\xi_3+\xi_T} + \frac{0.9\xi_2w^{-\frac{1}{2}}e^{-\frac{w}{2}}}{(0.95\xi_3 + \xi_T) \big(1-F_3(w)\big)} \nonumber \\
&< \frac{\xi_1\big(1-F_3(w) \big)}{0.95\xi_3+\xi_T} + \frac{0.9\xi_2}{0.95\xi_3+\xi_T}\sqrt{\frac{\pi}{2}}\cdot\frac{1}{w}
\end{align}
for all $w > w_T \geq 12$. The last line comes from substituting the following equation:
$$1-F_3(w) = \frac{1}{\sqrt{2\pi}} \int_w^{\infty}\sqrt{y}e^{-\frac{y}{2}}dy > \frac{\sqrt{w}}{\sqrt{2\pi}} \int_w^{\infty}e^{-\frac{y}{2}}dy = \sqrt{\frac{2w}{\pi} } e^{-\frac{w}{2}}$$

Lemma 1 in \cite{Laurent and Massart} gives $1-F_3(3+2\sqrt{3t}+2t) \leq e^{-t}$ for all $t>0$. Since $3+2\sqrt{3t} < 4t$ when $t \geq 2$, we have that $1-F_3(6t) < e^{-t}$ or $1-F_3(w) < e^{-\frac{w}{6}}$ for $w \geq 12$. Therefore, the rate of decay for $\epsilon_U/P_U$ is $\mathcal{O}(e^{-\frac{w}{6}}) + \mathcal{O}(\frac{1}{w}) $.

\end{document}